\documentclass[submitting,longbibliography,floatfix]{nst}
\usepackage{subfigure,dcolumn}
\usepackage{epstopdf}
\usepackage{chemformula}
\setchemformula{decimal-marker={,}}
\usepackage{upgreek}
\usepackage{tikz,xcolor,hyperref}
\definecolor{lime}{HTML}{A6CE39}
\DeclareRobustCommand{\orcidicon}{
	\begin{tikzpicture}
	\draw[lime, fill=lime] (0,0) 
	circle [radius=0.16] 
	node[white] {{\fontfamily{qag}\selectfont \tiny ID}};
	\draw[white, fill=white] (-0.0625,0.095) 
	circle [radius=0.007];
	\end{tikzpicture}
	\hspace{-2mm}
}
\foreach \x in {A, ..., Z}{%
	\expandafter\xdef\csname orcid\x\endcsname{\noexpand\href{https://orcid.org/\csname orcidauthor\x\endcsname}{\noexpand\orcidicon}}
}

\begin{document}
\title{Neutron skin and its effects in heavy-ion collisions}
\thanks{This work is partially supported by the National Natural Science Foundation of China 
(Nos. 11925502, 11935001, 12347106, 11961141003, 12147101 and 11890714), the Strategic Priority Research Program of Chinese Academy of Sciences (No. XDB34030000), the National Key R\&D Program of China (No. 2023YFA1606404).}

\author{Meng-Qi Ding\orcidA{}}
\affiliation{Key Laboratory of Nuclear Physics and Ion-beam Application (MOE), Institute of Modern Physics, Fudan University, Shanghai 200433, China}

\author{De-Qing Fang\orcidB{}}
\email{dqfang@fudan.edu.cn}
\affiliation{Key Laboratory of Nuclear Physics and Ion-beam Application (MOE), Institute of Modern Physics, Fudan University, Shanghai 200433, China}
\affiliation{Shanghai Research Center for Theoretical Nuclear Physics,
NSFC and Fudan University, Shanghai 200438, China}

\author{Yu-Gang Ma\orcidC{}}
\email{mayugang@fudan.edu.cn}
\affiliation{Key Laboratory of Nuclear Physics and Ion-beam Application (MOE), Institute of Modern Physics, Fudan University, Shanghai 200433, China}
\affiliation{Shanghai Research Center for Theoretical Nuclear Physics,
NSFC and Fudan University, Shanghai 200438, China}

\begin{abstract}
{Neutron skin is an exotic phenomena in unstable nuclei. The various effects in nuclear reactions caused by the neutron skin and also its relation with the properties of nuclear structure are reviewed in this article.
Based on numerous studies with theoretical models, strong correlations have been found between the neutron skin thickness and neutron removal cross section, neutron/proton yield ratio, t/\ch{^3He} yield ratio, neutron-proton momentum difference, isoscaling parameter, photon production, reaction cross sections for neutron induced reactions, charge-changing cross sections difference of mirror nuclei, astrophysical $S$-factor, and other quantities in nuclear reactions induced by neutron-rich nuclei.
Moreover, the relationships between neutron skin thickness and some properties of nuclear structure, such as $\alpha$-cluster formation, $\alpha$ decay, nuclear surface, nuclear temperature, and proton radii difference of mirror nuclei, have also been investigated. 
It also has been shown that the neutron skin plays a crucial role in relativistic heavy-ion collisions.
Experimentally, an unstable nucleus with neutron skin can be generated by radioactive nuclear beam facilities, and the thickness of neutron skin could be extracted by measuring the sensitive probes, which further helps giving stringent constraints on the equation of state of asymmetric nuclear matter and properties of neutron stars.}
\end{abstract}


\maketitle

\section{Introduction}\label{1}
The atomic nucleus is the microscopic structure of matter consisting of a certain number of protons and neutrons.
Theoretical prediction suggests that there may exists about 9000 nuclei~\cite{XIA20181, MA2021103911}.
Up to now, more than 3300 nuclei have been discovered, including 3340 nuclei in their ground state and 1938 excited isomers with half-life time longer than 100 ns~\cite{Kondev_2021}.
Except for the 339 naturally existing nuclei, the rest are artificially made. 
The number of naturally existing nuclei with isospin abundance data is only 289, which contains the 254 stable nuclei and the 35 long-lifetime radioactive ones~\cite{NPRchart}. 
Through abundant studies on stable nuclei, a relatively complete theory has established in traditional nuclear physics.
For example, the nuclear radius is proportional to $A^{1/3}$, where $A$ is the mass number.
And the density distributions of neutrons and protons are similar in a nucleus.
Additionally, nucleons are distributed in the orbitals with different energy levels according to the shell model. 
With the development of radioactive nuclear beam (RNB) experimental methods and the construction of a series of RNB facilities around the world, fast extension of nuclear chart towards the drip lines has revealed various new phenomena and physics in the domain of nuclear physics, such as the neutron halo, neutron skin, multi-nucleon clusters, shell evolution, exotic decaying modes, disappearance of normal magic numbers and appearance of new magic numbers~\cite{MA2021103911, Tanihata:2013jwa, NuPECC1352311, BLANK2008403, zhou2022NST, huo2022continuum,Fang-NT}. 

For the neutron-rich systems, due to the large asymmetry between proton ($p$) and neutron ($n$) numbers (denoting as $Z$ and $N$, respectively), these two kinds of fermions tend to be decoupled around the surface region of a nucleus, which is called the neutron skin.
The thickness of the neutron skin is defined as the difference between the root-mean-square (rms) radius of the neutron and that of the proton, i.e., $\langle r_{\rm n}^{2} \rangle ^{1/2}-\langle r_{\rm p}^{2}\rangle^{1/2}$.
The proton radius of a nucleus can be extracted with relatively high accuracy by electromagnetic interaction~\cite{ANGELI2004185}, but it is a challenge to obtain precise neutron radius directly in experiments due to the charge neutrality of neutrons.
So far, there are various experimental methods to measure the distribution of neutrons by strong or weak interaction probes, such as hadron scattering ~\cite{PhysRevLett.47.1436,PhysRevC.49.2118}, giant dipole resonance (GDR)~\cite{PhysRevLett.66.1287,SATCHLER1987215}, spin dipole resonance (SDR)~\cite{PhysRevLett.82.3216}, and antiprotonic annihilation~\cite{PhysRevC.57.2962,PhysRevLett.87.082501}.
However, the extracted results are usually model-dependent and vary greatly by different experimental analysis approaches~\cite{RAY1992223}.
Consequently, reliable probes for the neutron skin thickness are expected to be found which will have great significance in nuclear physics and astrophysics.

The equation of state (EOS) of nuclear matter is a fundamental problem in nuclear physics, which macroscopically reflects the microscopic nuclear interactions.
The EOS can be expressed as the binding energy per nucleon of nuclear matter
\begin{equation}\label{eq.1}
E(\rho,\delta) = E_0(\rho) + E_\text{sym}(\rho)\delta^2 + O(\delta^4),
\end{equation}
where $\rho_{n}$, $\rho_{p}$ and $\rho = \rho_{n} + \rho_{p}$ are the neutron, proton, and total densities, respectively, $\delta=(\rho_{n}-\rho_{p})/\rho$ is the isospin asymmetry, $E_0(\rho) = E(\rho, \delta=0)$ is the energy per nucleon of symmetric nuclear matter, and $E_\text{sym}(\rho)$ is the nuclear symmetry energy described by
\begin{equation}\label{eq.2}
E_\text{sym}(\rho) \equiv \left. \frac{1}{2!} \frac{\partial ^{2}E(\rho,\delta)}{\partial \delta^{2}}\right|_{\delta=0}.
\end{equation}
The properties of symmetric nuclear matter are relatively well-determined, while the isovector part remains largely uncertain and attracts widespread attention, which is helpful for a better understanding of the properties of radioactive nuclei, nuclear astrophysics, and dynamical evolution in heavy-ion collisions~\cite{LATTIMER2007109, BALDO2004241, ROCAMAZA201896, Jiang2017, Yu2020, Zhang2018, Wei:2021arw, BaoAnLiNPN1388681, LiLi:2022kvc, Liu:2021uoz}.
The density dependence of the symmetry energy is a crucial topic and shows great uncertainty when different nuclear forces or interaction potentials in theory are adopted, which can be regarded as soft or stiff~\cite{CLWEPJWeb}.
Around the saturation density $\rho_0$, $E_\text{sym}(\rho)$ can be expanded as
\begin{equation}\label{eq.3}
\begin{aligned}
E_\text{sym}(\rho) = &E_\text{sym}(\rho_0)+\frac{L}{3} \left( \frac{\rho-\rho_0}{\rho_0}\right) \\
&+ \frac{K_\text{sym}}{18}\left( \frac{\rho-\rho_0}{\rho_0}\right)^2+\cdots,
\end{aligned}
\end{equation}
where $L$ and $K_\text{sym}$ are the slope and curvature of the symmetry energy at the saturation density, respectively, defined by
\begin{equation}\label{eq.4}
L \equiv 3\rho_0\left. \frac{\partial E_\text{sym}(\rho)}{\partial \rho}\right|_{\rho=\rho_0},
\end{equation}
\begin{equation}\label{eq.5}
K_\text{sym} \equiv 9\rho_0^2\left. \frac{\partial ^{2}E_\text{sym}(\rho)}{\partial \rho^{2}}\right|_{\rho=\rho_0}.
\end{equation}
These two characteristic parameters play a critical role in determining the EOS of asymmetric nuclear matter because they govern the behavior of the symmetry energy at the subsaturation and oversaturation densities.
Based on various nuclear models, such as Skyrme, Gogny, and covariant models of different nature, it have been demonstrated that the neutron skin thickness is related to $E_\text{sym}(\rho_{0})$, $L$, $K_\text{sym}$, the ratio $L/E_\text{sym}(\rho_{0})$, $E_\text{sym}(\rho_{0})$--$a_{\text sym}$ (where $a_{\text sym}$ is the symmetry energy of finite nuclei), the ratio $E_\text{sym}(\rho_{0})$/$Q$ (where $Q$ is the surface stiffness coefficient), etc.~\cite{PhysRevC.72.064309, PhysRevLett.85.5296, PhysRevC.64.027302, FURNSTAHL200285, PhysRevC.65.044306, PhysRevLett.102.122502,HeWB-1}.
Moreover, Ref.~\cite{Zhang:2013wna} has proposed that the neutron skin thickness of heavy nuclei could be effectively determined by the EOS at a subsaturation cross density $\rho_{c} \approx 0.11$ fm$^{-3}$ rather than at $\rho_{0} \approx 0.16$ fm$^{-3}$.
As a result, a good knowledge of the neutron skin thickness will provide the possibility of obtaining information on the symmetry energy and determining the form of the EOS.

Furthermore, neutron stars are extremely dense compact stars, composed primarily of neutrons.
Despite differing by 18 orders of magnitude in spatial scale, the neutron skin and neutron stars have similar structures which are closely associated by the EOS of nuclear matter.
Detailed analyses have confirmed that some properties of neutron stars depend on the EOS so that the precise data of neutron skin thickness will help to constrain the specific properties of neutron stars.
Neutron stars, with a mass between that of a white dwarf and a black hole, are formed after a supernova explosion caused by gravitational collapse at the end of the evolution of a massive star.
A neutron star is one of the densest forms of matter in the universe, whose mass is about 1.5 times that of the sun and radius is about 12 km~\cite{science1090720}.
The density inside a neutron star can reach or even exceed 5--6 times that of normal nuclear matter which approximates that of the center of a heavy nucleus and is about $10^{14}$ times that of water.
It is generally accepted that the neutron stars consist of four major regions: the outer crust with nuclei and electron gas, the inner crust with nuclei, neutrons and electrons, the outer core with uniformly-distributed electrically-neutral nuclear matter, and the dense inner core probably containing other particles such as hyperons and quarks.
The first evidence for the existence of neutron stars came with the discovery of pulsars in 1967.
The detection of gravitational waves from a binary neutron star merger (GW170817) by the LIGO-Virgo Collaboration~\cite{PhysRevLett.119.161101} opens the new era of multimessenger astronomy and illuminates the importance of astrophysics in exploring the nature of dense matter and the synthesis of heavy elements.
There is no direct connection between the neutron skin and neutron stars, but they are linked by the EOS of nuclear matter.
The slope of the symmetry energy corresponds to the pressure of nuclear matter which is very closely related to both the neutron skin thickness and the radii of neutron stars.
The relation between the neutron skin thickness and neutron-star properties helps us to understand the EOS of nuclear matter more comprehensively, and also gives us opportunities to study the properties of neutron stars in the laboratory.
The dependences of the neutron skin thickness or the EOS on neutron-star properties have been studied extensively, such as the mass-radius relation~\cite{PhysRevD.82.101301, Steiner_2010, Carriere_2003}, the crust-core transition~\cite{PhysRevLett.86.5647, PhysRevLett.83.3362, Lattimer_2001, Xie2023BayesianIO}, the onset of the direct Urca process~\cite{Page_2004, PhysRevC.66.055803}, and the stellar moment of inertia~\cite{PhysRevC.82.025810, STEINER2005325}.
For example, the $K_\text{sym}$--$L$ correlation has a great influence on the crust-core transition, radius, and tidal deformability of canonical neutron stars, especially at small $L$ values~\cite{PhysRevC.102.045807}.
And with a variety of interactions or theoretical models, different EOS correspond to different mass-radius relation of the neutron star~\cite{LATTIMER2016127}.
In addition, Ref.~\cite{PhysRevLett.120.172702} provided an upper limit on the neutron skin thickness of \ch{^{208}Pb} based on the GW170817 data, and conversely provided a lower limit on the tidal polarizability by relying on the lower limit on the measured neutron skin thickness of \ch{^{208}Pb} via parity violating electron nucleus scattering, which can further infer the properties of the EOS~\cite{PhysRevLett.120.172702}.
And Ref.~\cite{PhysRevC.101.034303} discussed the influence of neutron skin, isospin diffusion, and tidal deformability and maximum mass of a neutron star on the constraints of the symmetry energy and its associated nuclear matter parameters.
While the correlations between the neutron skin thickness and several neutron-star observables have been proved to exist by a variety of studies, these results have some system dependence and uncertainty.
In order to quantitatively describe the degree of correlations between various observables, Ref.~\cite{PhysRevC.86.015802} developed a covariance analysis to quantify the theoretical errors and correlation coefficients.
Recently, statistical analysis has been carried out from the cores of neutron stars to the crust. 
For example, by using the pure neutron matter and neutron skin data, a Bayesian inference of neutron-star crust properties is presented in Ref.~\cite{NEWTON2022137481} based on a compressible liquid drop model with an extended Skyrme EDF, such as the crust-core transition pressure, the crust-core transition baryon chemical potential, and the thickness of the pasta phases relative to the crust.

A comprehensive review of the neutron skin's effects in nuclear reactions is given in this article.
In Sec.~\ref{2}, the previous studies on the effects of neutron skin in nuclear reactions by diverse theoretical or phenomenological approaches are briefly presented. In Sec.~~\ref{3}, the recent progresses on neutron skin effects are presented in the studies in relativistic heavy-ion collisions. The experimental detection methods for the neutron skin are introduced in Sec.~\ref{4}.
Finally, a summary is given in Sec.\ref{5}.

\section{Effects of neutron skin in nuclear reactions}\label{2}
Compared with stable nuclei, the nuclei with neutron skin exhibit unique structural characteristics and their density distributions of neutrons are more extended than those of protons, which lead to all kinds of effects in nuclear reactions.
In this review, the correlations between the neutron skin and several quantities are introduced: (A) neutron removal cross section, (B) yield ratios of light particles (neutron/proton, triton/\ch{^3He}), (C) momentum difference between neutrons and protons, (D) projectile fragmentation, (E) photon emission, (F) reaction cross sections for nucleon induced reactions, (G) properties of nuclear structure ($\alpha$-cluster formation, $\alpha$-decay half-life, nuclear surface, nuclear temperature), (H) properties of mirror nuclei (proton rms radii difference of mirror nuclei, charge-changing cross sections difference of mirror nuclei), and (I) astrophysical $S$-factor.

\subsection{Neutron skin and neutron removal cross section} 
\begin{figure}[htbp]
\includegraphics[width=0.85\columnwidth]{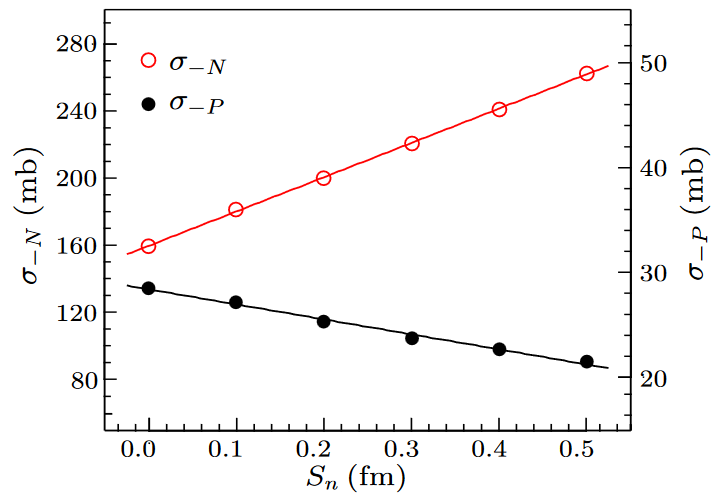}
\caption{Correlation between the neutron removal cross section (left Y axis), proton removal cross section (right Y axis) and neutron skin thickness for \ch{^{48}Ca} + \ch{^{12}C} reactions at 100 MeV/nucleon.~\cite{Fang_2011}}
\label{fig.1}
\end{figure}

The statistical abrasion ablation (SAA) model was developed to describe heavy-ion collisions, considering nuclear reactions as two-step process, that is, abrasion and evaporation.
This model treats the neutron and proton density distributions separately and makes complete statistics of nucleon-nucleon collisions in the overlapping region of the two colliding nuclei, which can well describe the fluctuation of the neutron-to-proton ratio in the process of fragmentation and study the isospin effect in nuclear reactions~\cite{BROHM1994821}.
Based on the SAA model, the relation between the neutron skin thickness and neutron removal cross section ($\sigma_{-N}$) is studied in Refs.~\cite{PhysRevC.81.047603,Fang_2011}.
Different sizes of neutron skin are obtained by adjusting the diffuseness parameter of neutrons in the two-parameter Fermi distribution.
And the neutron removal cross section is defined as the summation of production cross sections for fragments with the same proton number as the projectile but with a smaller neutron number.
It is confirmed that there is a strong correlation between the neutron skin thickness and neutron removal cross section~\cite{PhysRevC.81.047603}.
Similarly, the proton removal cross section ($\sigma_{-P}$) is defined as the summation of production cross sections for fragments with the same neutron number as the projectile.
As illustrated in Fig.~\ref{fig.1}, with the increase of the neutron skin thickness, the proton removal cross section decreases, which displays the opposite dependence compared to the neutron removal cross section~\cite{Fang_2011}.
It reveals that for the nuclei with large neutron skin, the neutrons will have a more expanded spatial distribution than protons so that the neutrons may be easier to be removed in the peripheral collisions.
Furthermore, in order to eliminate the systematic measurement error in experiments, the ratio between the (one-)neutron and proton removal cross section is also explored, which is found to be proportional to the neutron skin thickness and be a more sensitive probe for the determination of the neutron skin thickness.
Moreover, Ref.~\cite{PhysRevLett.119.262501} has also demonstrated that the neutron removal cross section is sensitive to the neutron skin thickness for Sn isotopes calculated by the relativistic mean-field (RMF) theory based on variations of the DD2 interaction, which potentially provides constraints on the symmetry energy.

\subsection{Neutron skin and yield ratios of light particles} 
\begin{figure}[htbp]
\includegraphics[width=0.85\columnwidth]{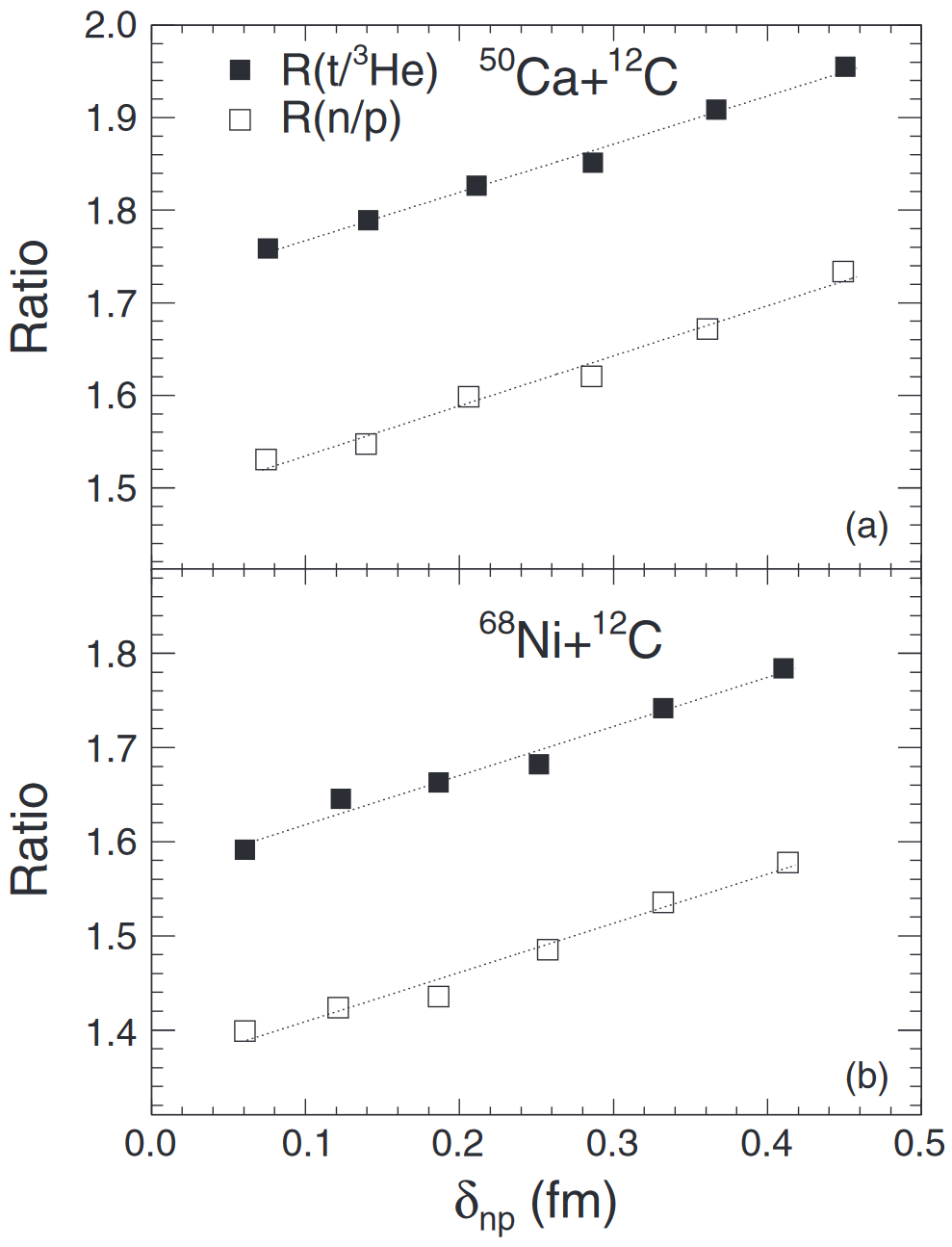}
\caption{Correlation between the neutron-to-proton yield ratio, triton-to-\ch{^3He} yield ratio, and neutron skin thickness for \ch{^{50}Ca} + \ch{^{12}C} (a) and \ch{^{68}Ni} + \ch{^{12}C} (b) at 50 MeV/nucleon.~\cite{PhysRevC.89.014613}}
\label{fig.2}
\end{figure}

The isospin-dependent quantum molecular dynamics (IQMD) model is a semi-classical microscopic transport theory developed from the traditional QMD model with the appropriate consideration of isospin degrees of freedom, which provides reasonable explanations for the nuclear reactions induced by nuclei far from the $\beta$-stability line~\cite{AICHELIN1991233, Ying-XunZhang, TMEP:2022xjg, Colonna:2021xuh, Aichelin:1988me, Feng:2018emx, PhysRevC.58.2283, PhysRevC.86.044620, LIU200524, PhysRevC.87.014621, PhysRevC.51.710, PhysRevC.60.064604, PhysRevC.73.014604, PhysRevC.51.3256}.
In this model, each nucleon is regarded as a Gaussian wave packet with finite width instead of a classical point particle, and three basic components of dynamics in intermediate-energy heavy-ion collisions are involved: the mean field, two-body collisions, and Pauli blocking.
Within the framework of the IQMD model, the collision processes of Ca and Ni isotopes with \ch{^{12}C} target at the incident energy of 50 MeV/nucleon are simulated to investigate the dependence between the neutron skin thickness and the yield ratio of the emitted neutrons and protons [$R(n/p)$] in Refs.~\cite{SUN2010396,SunXY2011A}.
The neutron and proton density distributions used for the phase-space initialization in the IQMD model come from the Droplet model, which can gain different values of the neutron skin thickness by changing the diffuseness parameter of the neutron density for the neutron-rich projectile.
With different impact parameters, it is found that from central to peripheral collisions the neutron skin thickness is always proportional to $R(n/p)$, which is most sensitive in peripheral collisions as the difference of density distributions between neutrons and protons is mainly reflected in the surface region of the neutron-rich nuclei.
And Ref.~\cite{PhysRevC.109.024616} suggested that the effects of the neutron skin thickness on $R(n/p)$ are more prominent at lower momentum.
Therefore, $R(n/p)$ could be viewed as an experimental observable to extract the neutron skin thickness. 
Nevertheless, it brings some challenge to measure $R(n/p)$ precisely in experiments because of the low detection efficiency for neutrons.
Hence Ref.~\cite{PhysRevC.89.014613} proposed to explore the relatively heavier charged particles, such as triton and \ch{^3He}, which are much easier to be measured experimentally.
It can be seen from Fig.~\ref{fig.2} that there is a linear correlation between the neutron skin thickness and the yield ratio of triton to \ch{^3He} [$R(t/\ch{^3He})$].
And for different neutron skin thicknesses, $R(t/\ch{^3He})$ is proportional to $R(n/p)$, that is to say, the double ratio $R(t/\ch{^3He})/R(n/p)$ is almost constant.
In consequence, both $R(n/p)$ and $R(t/\ch{^3He})$ could be used as possible experimental probes for the neutron skin thickness, which has also been shown by Refs.~\cite{PhysRevC.101.054601, Yan:2019ult, PhysRevC.109.024616} subsequently.
However, $R(t/\ch{^3He})$ could give higher precision in extracting the neutron skin thickness because the yields of charged particles can be measured quite accurately.

\subsection{Neutron skin and the difference of momentum between neutrons and protons} 
\begin{figure}[htbp]
\includegraphics[width=0.9\columnwidth]{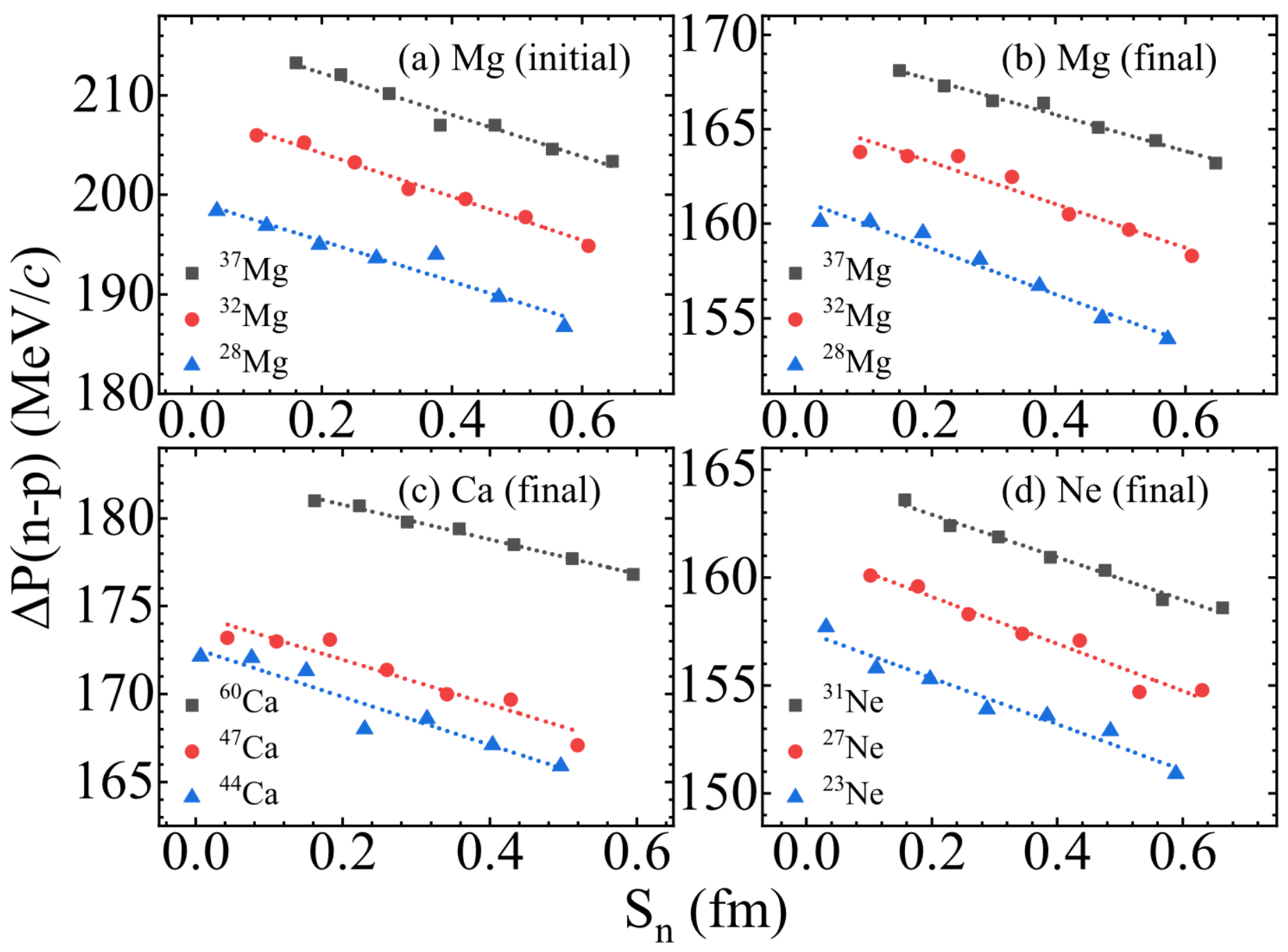}
\caption{Correlation between the neutron-proton momentum difference and neutron skin thickness for \ch{^{28,32,37}Mg} + \ch{^{12}C} in the initial state (a), \ch{^{28,32,37}Mg} + \ch{^{12}C} in the final state (b), \ch{^{44,47,60}Ca} + \ch{^{12}C} in the final state (c), and \ch{^{23,27,31}Ne} + \ch{^{12}C} in the final state (d) at 50 MeV/nucleon.~\cite{PhysRevC.109.024616}}
\label{fig.3}
\end{figure}

In terms of the wave functions in quantum mechanics, the density distribution and momentum distribution are each other's Fourier transforms.
And according to the uncertainty principle, the smaller intrinsic momentum fluctuations between fragments mean more dispersion in coordinate space.
It is easier to measure the momentum than the density distribution for neutrons in experiments.
Therefore, with the help of the IQMD model, Ref.~\cite{PhysRevC.109.024616} studied the dependence of the difference of momentum between neutrons and protons with the neutron skin thickness by simulating the semiperipheral collisions of neutron-rich Ca, Mg and Ne isotopes on \ch{^{12}C} target at 50 MeV/nucleon.
The difference of momentum between neutrons and protons is defined as the average magnitude of the relative momentum in every possible permutation and combination.
It is found that the neutron-proton momentum difference decreases linearly with the increase of the neutron skin thickness in both initial and final states as illustrated in Fig.~\ref{fig.3}.
It means that if neutrons and protons get closer in space distribution, they will have larger difference in momentum distribution.
Further study shows that the normalized momentum spectra of the emitted neutrons are evidently distinct for different neutron skin thicknesses, while those of protons are almost the same.
And with the thicker neutron skin, the peak position of the momentum spectrum of neutrons is smaller, indicating the average momentum of the emitted neutrons becomes smaller.
Furthermore, the difference of momentum between tritons and \ch{^3He} can be taken as a probe for the neutron skin thickness of the nucleus with a large isospin asymmetry, such as \ch{^{60}Ca}, \ch{^{37}Mg}, and \ch{^{31}Ne}.
With the development of the detector technology, the detection efficiency for neutrons has been improved, which makes it possible to extract useful information on the neutron skin thickness from the neutron-proton momentum difference in future experiments.

\subsection{Neutron skin and projectile fragmentation} 
The projectile fragmentation is a well-established technique to produce rare isotopes by bombarding the projectile nucleus on the target nucleus with a certain incident energy, which can produce varieties of fragments with a wide mass region from nucleons to heavy fragments and a wide isospin region from neutron-deficient to neutron-rich isotopes~\cite{PhysRevC.98.014610, Ma_2022, wei2022NST}.
This process can be well explained by the participant-spectator model~\cite{osti_5922662}.
In the peripheral collision, the overlap zone between the projectile and target (called the participant) generates the abrasion and evaporation of nucleons or light fragments, while the residues with the charge number close to the projectile (called the spectator or projectile-like fragments) fly away at a nearly unchanged velocity, which may retain much information about the initial density distribution of the projectile.
Therefore, some observables in the projectile fragmentation have been studied for their relevance with the neutron skin, which is conducive to extract useful information on the nuclear structure and the EOS of asymmetric nuclear matter, and also discuss the reaction mechanism of nuclear collisions.

\begin{figure*}[htbp]
\includegraphics[width=0.95\textwidth]{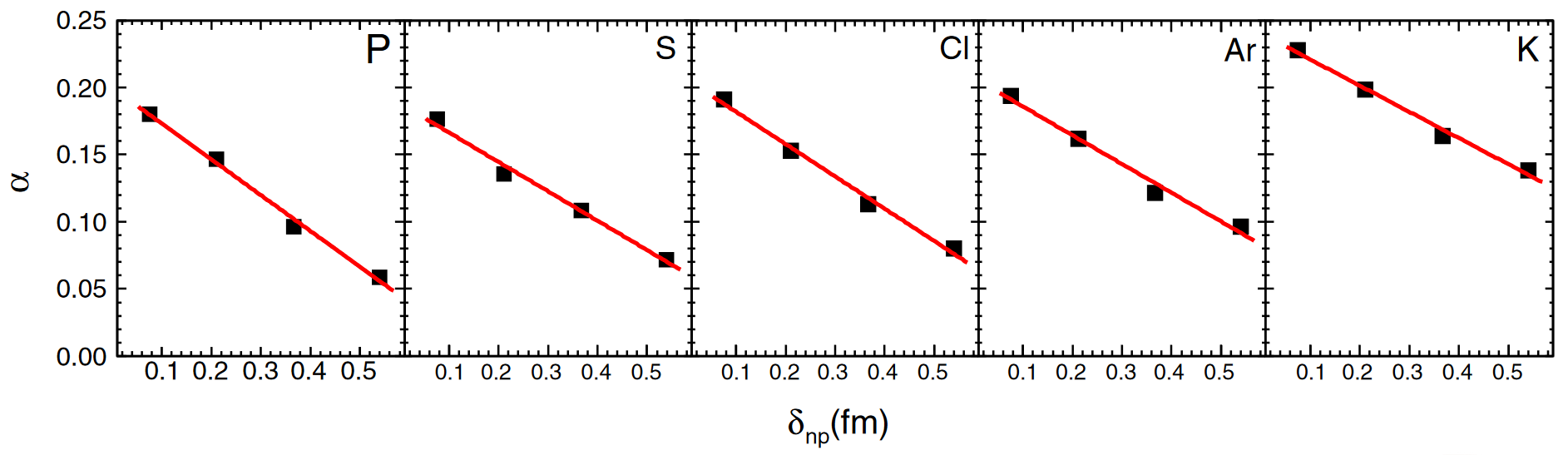}
\caption{Correlation between the isoscaling parameter $\alpha$ and neutron skin thickness for fragments with the proton number varying from 15 to 19 based on the two reactions \ch{^{50}Ca} + \ch{^{12}C} and \ch{^{48}Ca} + \ch{^{12}C} at 50 MeV/nucleon.~\cite{PhysRevC.91.034618}}
\label{fig.4}
\end{figure*}

The isoscaling phenomenon is a fascinating topic in the survey of isospin effects in fragmentation dynamics, which has been observed in both theoretical and experimental studies.
Based on two similar nuclear reactions that differ only in the isospin asymmetry, it is found that the yield ratio of a given isotope in the two reactions (denoted as 1 and 2, respectively) obeys a scaling law and have exponential dependencies on the neutron number and proton number, which can expressed as
\begin{equation}\label{eq.6}
R_{21}(N,Z) = Y_{2}(N,Z)/Y_{1}(N,Z) = C\exp(\alpha N+\beta Z),
\end{equation}
where $\alpha$ and $\beta$ are isoscaling parameters, and $C$ is a normalization constant~\cite{BARAN2005335}.
In the grand-canonical approximation, $\alpha$ and $\beta$ are equal to the difference of the chemical potentials between the two reaction systems for neutron and proton, respectively.
Systematical isoscaling analyses have been carried out on the light fragments with $Z \leq 8$, which demonstrates that the isoscaling parameters keep constant basically for different fragments and have a linear dependence on the symmetry energy, making it feasible to indirectly determine the symmetry energy term of the EOS.
As a result, the isoscaling behavior of heavy fragments would also be of considerable interest.
In Ref.~\cite{PhysRevC.91.034618}, the sensitivity of isoscaling behavior to the neutron skin thickness for projectile-like fragments is investigated by simulating the peripheral collisions of \ch{^{50}Ca} + \ch{^{12}C} and \ch{^{48}Ca} + \ch{^{12}C} at 50 MeV/nucleon through the IQMD model followed by the GEMINI decay code.
It can be seen from Fig.~\ref{fig.4} that with the increase of the neutron skin thickness of \ch{^{50}Ca}, the extracted isoscaling parameter $\alpha$ decreases linearly.
Besides, under the same neutron skin thickness, $\alpha$ is not an constant for the heavy fragments with different proton numbers, which exhibits different isoscaling behaviour from light particles, probably arising from the different formation mechanisms.
Moreover, the average value of $N/Z$ of the projectile-like fragments also appears a negative linear relationship with the neutron skin thickness.
As a consequence, the experimental measurements of both the isoscaling parameter $\alpha$ and mean $N/Z$ of projectile-like fragments could provide a valuable evaluation of the neutron skin thickness to further set stringent constraints on the critical parameters of the EOS.
However, the distortion to isoscaling phenomenon is found in Ref.\cite{Peng2024}.
Based on the experimental data and two different theoretical models, they examined the isoscaling properties for neutron-rich fragments produced in highly asymmetric systems, namely \ch{^{40,48}Ca} + \ch{^{9}Be} and \ch{^{58,64}Ni} + \ch{^{9}Be} at 140 MeV/nucleon.
They proposed that the isoscaling law is obeyed when the fragments have the same range of the neutron excess $I=N-Z$, which means the fragments share the same environment, i.e., the same colliding region, temperature, and nuclear density.
Additionally, the correlation between $|\beta|$ and $\alpha$ varies for the fragments with different $I$ values, indicating that the $|\beta|/\alpha$ ratio for a specific fragment can be adopted to detect the differences in neutron and proton densities in different regions of the nucleus.

\begin{figure}[htbp]
\includegraphics[width=0.75\columnwidth]{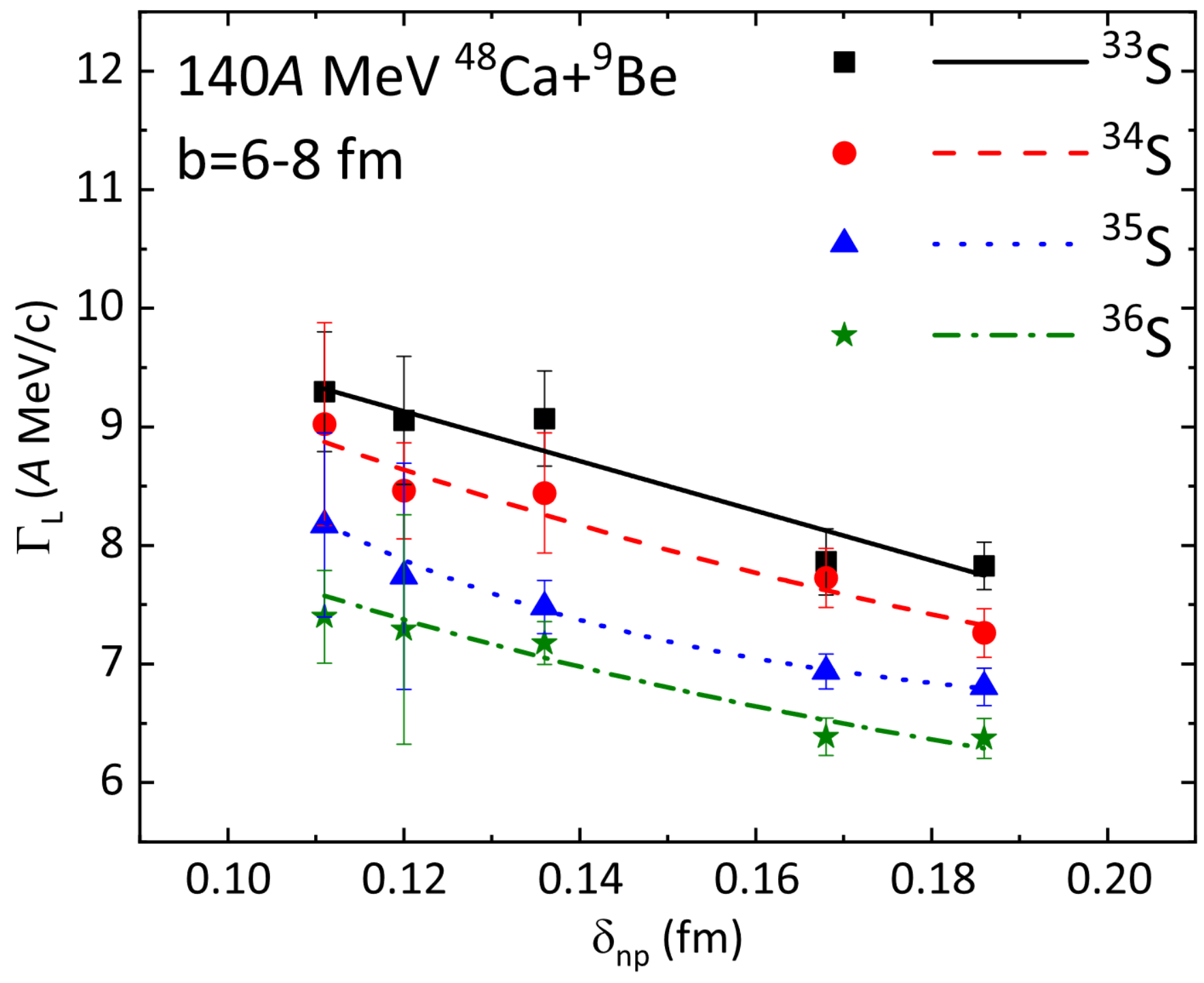}
\caption{Correlation between the width of the left side of the parallel momentum distribution for neutron-rich sulfur fragments and neutron skin thickness of \ch{^{48}Ca} for \ch{^{48}Ca} + \ch{^{9}Be} at 140 MeV/nucleon.~\cite{CWMa2024}}
\label{fig.5}
\end{figure}

Recently, the parallel momentum distribution ($p_{\|}$) of the residual fragments was studied in Ref.\cite{CWMa2024} by simulating the \ch{^{48}Ca} + \ch{^{9}Be} projectile fragmentation reaction at 140 MeV/nucleon within the framework of the Lanzhou Quantum Molecular Dynamics (LQMD) model which is an isospin- and momentum-dependent transport model.
The initial density distributions adopt the Fermi distribution with two parameters.
The $p_{\|}$ is nonsymmetric and can be fitted by a combined Gaussian function, resulting in different width parameters for the left side ($\Gamma_L$) and the right side ($\Gamma_R$) of the distribution.
It is confirmed that the $\Gamma_L$ of the projectile-like fragments in peripheral collisions is sensitive to the neutron skin thickness of the neutron-rich projectile nucleus as shown in Fig~\ref{fig.5}.
The $p_{\|}$ is easily measured in experiments so that the $\Gamma_L$ of the $p_{\|}$ can serve as a possible probe to the neutron skin thickness.

\begin{figure}[htbp]
\includegraphics[width=0.95\columnwidth]{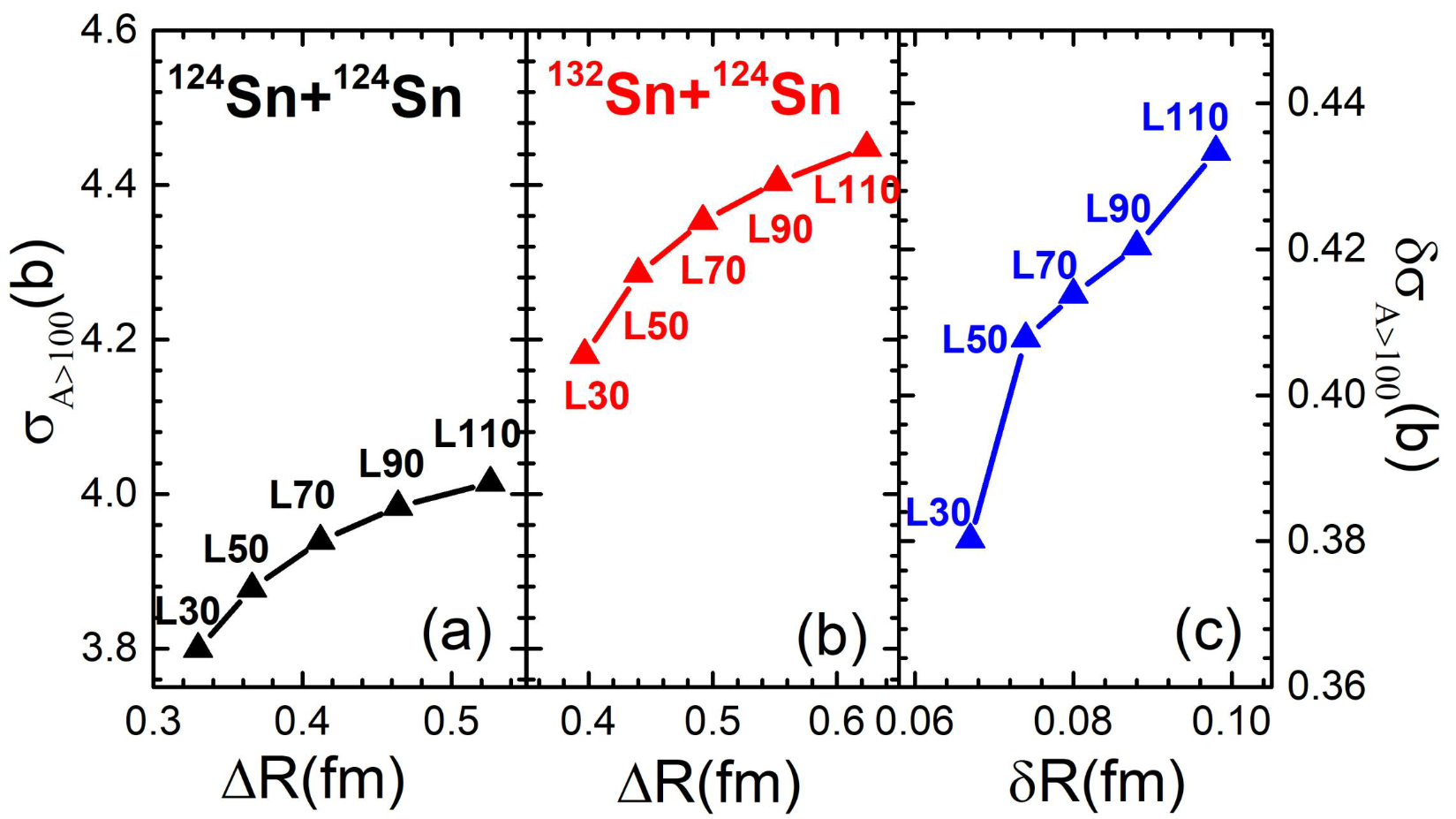}
\caption{Correlation between the cross sections of the projectile-like residues with $A>100$ and neutron skin thickness for \ch{^{124}Sn} + \ch{^{124}Sn} (a) and \ch{^{132}Sn} + \ch{^{124}Sn} (b) at 200 MeV/nucleon.
Correlation between the difference of $\sigma_{A>100}$ and difference of neutron skin thickness between the two systems (c).~\cite{universe9050206}}
\label{fig.6}
\end{figure}

Furthermore, the effect of the neutron skin on the cross sections of the primary projectile-like residues is presented in Ref.~\cite{universe9050206}.
The collisions of \ch{^{124, 132}Sn} + \ch{^{124}Sn} at 200 MeV/nucleon are calculated by developing a new version of an improved quantum molecular dynamics (ImQMD-L) model, in which the neutron skin of nuclei in the initialization and the mean-field potential in nucleon propagation are consistently treated.
And the thickness of neutron skin is altered by employing five sets of Skyrme parameters, corresponding to different values of the slope of the symmetry energy varying from 30 to 110 MeV.
As shown in Fig.~\ref{fig.6}(a) and (b), the cross sections of the primary projectile-like residues with mass number $A>100$ ($\sigma_{A>100}$) are positively correlated with the neutron skin thickness and the slope of the symmetry energy for both systems.
For the sake of suppressing the systematic uncertainty caused by the model, the difference of $\sigma_{A>100}$ between the two systems ($\delta \sigma_{A>100}$) is further investigated, which confirms that $\delta \sigma_{A>100}$ is still associated with the difference of the neutron skin thickness between the two systems and the slope of the symmetry energy (see Fig.~\ref{fig.6}(c)).
Thus it is potential to take $\sigma_{A>100}$ or $\delta \sigma_{A>100}$ as the experimental probes for determining the neutron skin thickness of the unstable nuclei and constraining the symmetry energy.

\begin{figure}[htbp]
\includegraphics[width=0.75\columnwidth]{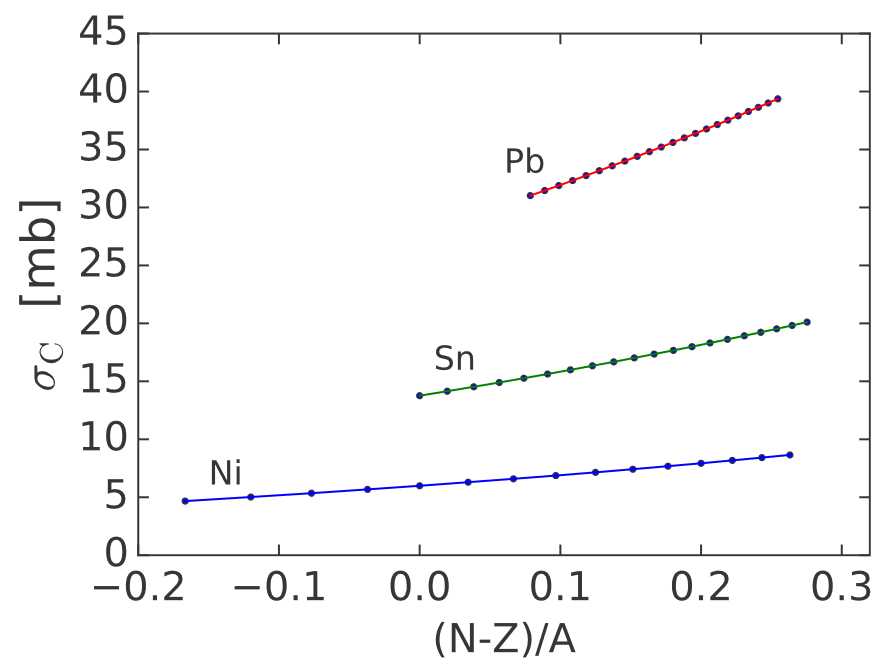}
\caption{Correlation between the charge-changing cross sections and isospin asymmetry of the projectiles for Ni, Sn, and Pb isotopes on C targets at 1 GeV/nucleon.~\cite{PhysRevC.100.015802}}
\label{fig.7}
\end{figure}

Some other observables in fragmentation reactions are studied in Ref.~\cite{PhysRevC.100.015802}, including neutron removal cross sections, charge-changing cross sections, and total interaction cross sections.
It indicates that these quantities are dependent on the neutron skin thickness, which could be used to give some constraints on the slope of the symmetry energy and have a better understanding of the properties of neutron stars.
For example, Fig.~\ref{fig.7} illustrates that the charge-changing cross sections tend to increase linearly with the rising isospin asymmetry, namely, the larger neutron skin sizes of the projectile.
Additionally, configurational information entropy (CIE) analysis has been adopted to conclude that the related CIE quantities, e.g., the isotopic/mass/charge cross section distributions, in the projectile fragmentation reactions are sensitive to the neutron skin thickness of neutron-rich nuclei~\cite{Wei:2022iuy,Ma:2022lox}.

\subsection{Neutron skin and photon emission}
Compared with the conventional hadronic probes, 
photons interact weakly with the nuclear medium and only by the electromagnetic interaction~\cite{PhysRevC.85.024618, Deng:2018frf}. 
Hence they are not interfered by the final-state interactions, so that a more realistic picture of nuclear matter could be obtained.
Both theoretical and experimental results have demonstrated that the hard photons, i.e., photons with energies larger than 30 MeV, primarily originate from the incoherent proton-neutron bremsstrahlung collisions in intermediate-energy heavy-ion collisions. Those photons are emitted from two distinct sources, that is, direct hard photons and thermal hard photons.
The direct hard photons derive from the earlier stage of the reaction which may reserve the memory of the initial projectile.
In Ref.~\cite{PhysRevC.105.034616}, the effects of neutron skin on direct hard photon emission are explored from the reactions of \ch{^{50}Ca} + \ch{^{12}C} and \ch{^{50}Ca} + \ch{^{40}Ca} within the framework of the IQMD model, in which the channel of incoherent proton-neutron bremsstrahlung collisions is embedded.
It is found that more direct hard photons are produced in peripheral collisions for thicker neutron skin.
Moreover, it is confirmed that the yield ratio of direct hard photons between central and peripheral collisions in the same reaction system [$R_{cp}(\sigma_{\gamma})$] shows a decreasing trend with the increase of the neutron skin thickness, which is more sensitive at higher incident energies.
Meanwhile, it is concluded that the neutron skin has a great effect on the rapidity dependence of multiplicity ($N_{\gamma}$) and multiplicity ratio between central and peripheral collisions [$R_{cp}(N_{\gamma})$] for direct hard photons.
With the increase of the neutron skin thickness, $N_{\gamma}$ tends to increase, while $R_{cp}(N_{\gamma})$ decreases.
Besides, the rapidity dependence of $R_{cp}(N_{\gamma})$ exhibits different behaviours for the two reaction systems.
By adjusting the neutron skin thickness of the projectile \ch{^{50}Ca}, the biggest difference of the rapidity distribution of $R_{cp}(N_{\gamma})$ appears near the target nucleus side for the target \ch{^{12}C}, while at around zero rapidity for the target \ch{^{40}Ca} which is a more symmetric system.
Therefore, direct hard photon emission could be used as an experimental observable to extract information on the neutron skin thickness and is much cleaner in comparison with the above probes, such as nucleons, light fragments, and projectile-like fragments produced in the reaction.
References~\cite{PhysRevC.92.014614, PhysRevC.108.034617} have also demonstrated that the production of hard photons is sensitive to the neutron skin thickness.
Further theoretical researches are necessary to explore the dependence of direct hard photon emission on the EOS and symmetry energy.

\subsection{Neutron skin and reaction cross sections for nucleon induced reactions}
The isospin effects in proton (neutron) induced reactions on Sn isotopes at 100 MeV are investigated in Ref.~\cite{OuLi2009} by means of the ImQMD model.
The reaction cross sections for nucleon induced reactions are influenced by both the neutron skin thickness of the target nuclei combined with the isospin dependence of the nucleon-nucleon cross sections and the motion of the incident nucleon from the nuclear mean field.
The effects of density dependence of symmetry energy in nucleon induced reactions have been systematically studied in Ref.~\cite{PhysRevC.78.044609} so Ref.~\cite{OuLi2009} further attempts to disentangle the effects of the neutron skin thickness and symmetry potential.
They artificially make two sets of initial nuclei which not only have different neutron skin thicknesses but also have correct binding energies.
It indicates that the reaction cross sections for proton induced reactions are sensitive to the density dependence of the symmetry energy but less sensitive to the neutron skin thickness of the target nuclei, while the neutron induced reactions exhibit opposite effects.
The reaction cross sections for neutron induced reactions are less sensitive to the density dependence of the symmetry energy but sensitive to the neutron skin thickness of the target nuclei, which potentially provides an approach to  obtaining information on the neutron skin thickness.

\subsection{Neutron skin and properties of nuclear structure}
\begin{figure}[htbp]
\includegraphics[width=0.95\columnwidth]{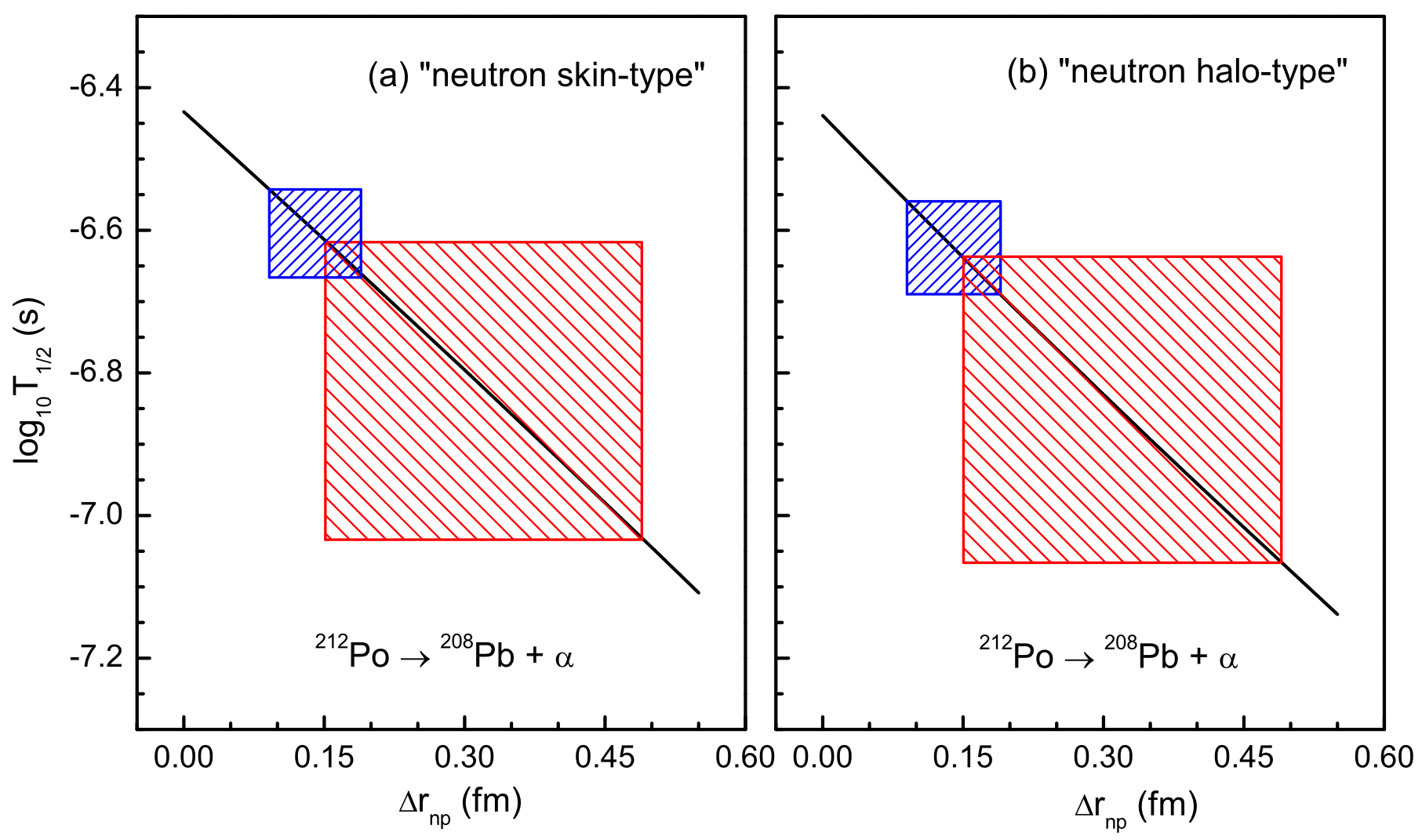}
\caption{Correlation between the theoretical $\alpha$-decay half-life of the parent nucleus \ch{^{212}Po} and neutron skin thickness of the daughter nucleus \ch{^{208}Pb} with “neutron skin-type” distribution (a) and “neutron halo-type” distribution (b).
The red and blue zones correspond to the neutron skin thickness deduced from parity-violating electron scattering~\cite{PhysRevLett.108.112502} and coherent pion photoproduction cross sections~\cite{PhysRevLett.112.242502}, respectively.~\cite{PhysRevC.92.054322}}
\label{fig.8}
\end{figure}

Owing to the structural particularity of the neutron skin, several characteristics of nuclear structure may be potential probes of the neutron skin.
Theoretical studies have suggested that the formation of $\alpha$ clusters in heavy nuclei is closely related with the properties of $\alpha$ decay~\cite{QI2019214, Gamow1928ZurQD} and has a certain probability of occurring at the very surface of the nucleus in the ground state~\cite{PhysRevC.81.015803, PhysRevC.87.041302, PhysRevLett.108.062702}.
The surface $\alpha$-clustering phenomenon also appears in neutron-rich heavy nuclei, leading to a tight connection between the $\alpha$-cluster formation and neutron skin thickness. It implies that the growth of neutron skin will suppress the $\alpha$ cluster at nuclear surface~\cite{PhysRevC.81.015803, PhysRevC.89.064321, RevModPhys.89.015007}.
It has been confirmed experimentally by the monotonous reduction of proton-induced $\alpha$-knockout reaction $(p, p\alpha)$ cross section with increasing mass number for neutron-rich Sn isotopes~\cite{doi:10.1126/science.abe4688}.
Recently, the negative correlation between the neutron skin thickness and $\alpha$ clustering in C isotopes is further verified within the framework of antisymmetrized molecular dynamics (AMD)~\cite{ZHAOEPJA210504}. 
In addition, in the $\alpha$ decay of heavy nuclei, the influences of the neutron skin thickness on the $\alpha$-decay half-life are explored in Refs.~\cite{Wan:2016inr, PhysRevC.96.054328, PhysRevC.90.064310, PhysRevC.92.054322}.
For instance, by using the generalized density-dependent cluster model, Ref.~\cite{PhysRevC.92.054322} suggests that there is a negative linear correlation between the calculated $\alpha$-decay half-life and the neutron skin thickness of the daughter nuclei as displayed in Fig.~\ref{fig.8}, which implies that the increase of the neutron skin thickness would hinder $\alpha$ decay. 

It has been theoretically expected that the difference of the proton and neutron surface widths is of great significance in determining the neutron skin thickness~\cite{PhysRevC.96.035804}.
An appropriate correction of the nuclear surface diffuseness to the neutron skin thickness was applied in the macroscopic model, i.e., a compressible droplet model where the nuclear density distribution follows the Fermi distribution with two parameters.
The deduced macroscopical neutron skin thicknesses are compared with the results calculated by a microscopic Skyrme-Hartree-Fock(SHF)+BCS method with several sets of Skyrme effective interactions, which confirms that there exists obvious dependence between the surface widths and the neutron skin thickness.

The temperature dependence of the neutron skin is analyzed in Ref.~\cite{Yuksel:2014gra} by using the finite temperature SHF+BCS approximation.
It has been known that at the critical temperature, the pairing correlations disappear and the phase change from superfluid to the normal state happens.
The study discovers that the proton radius of \ch{^{120}Sn} remains constant up to the critical temperature, while the neutron radius falls a little.
However, there is a rising trend in both proton and neutron radius above the critical temperature.
Hence with the elevation of temperature, the neutron skin thickness firstly decreases slightly and then increases substantially, which has the minimum value at the critical temperature.
The effect of $N/Z$ value on the temperature dependence of the neutron skin is further investigated in Sn isotopic chain. It is demonstrated that the increase in the proton radius hinders the formation of the neutron skin in less-neutron rich nuclei.
These results may be caused by the impact of temperature on the occupation probabilities of the single-particle states around the Fermi level. 

\subsection{Neutron skin and properties of mirror nuclei}
\begin{figure}[htbp]
\includegraphics[width=0.85\columnwidth]{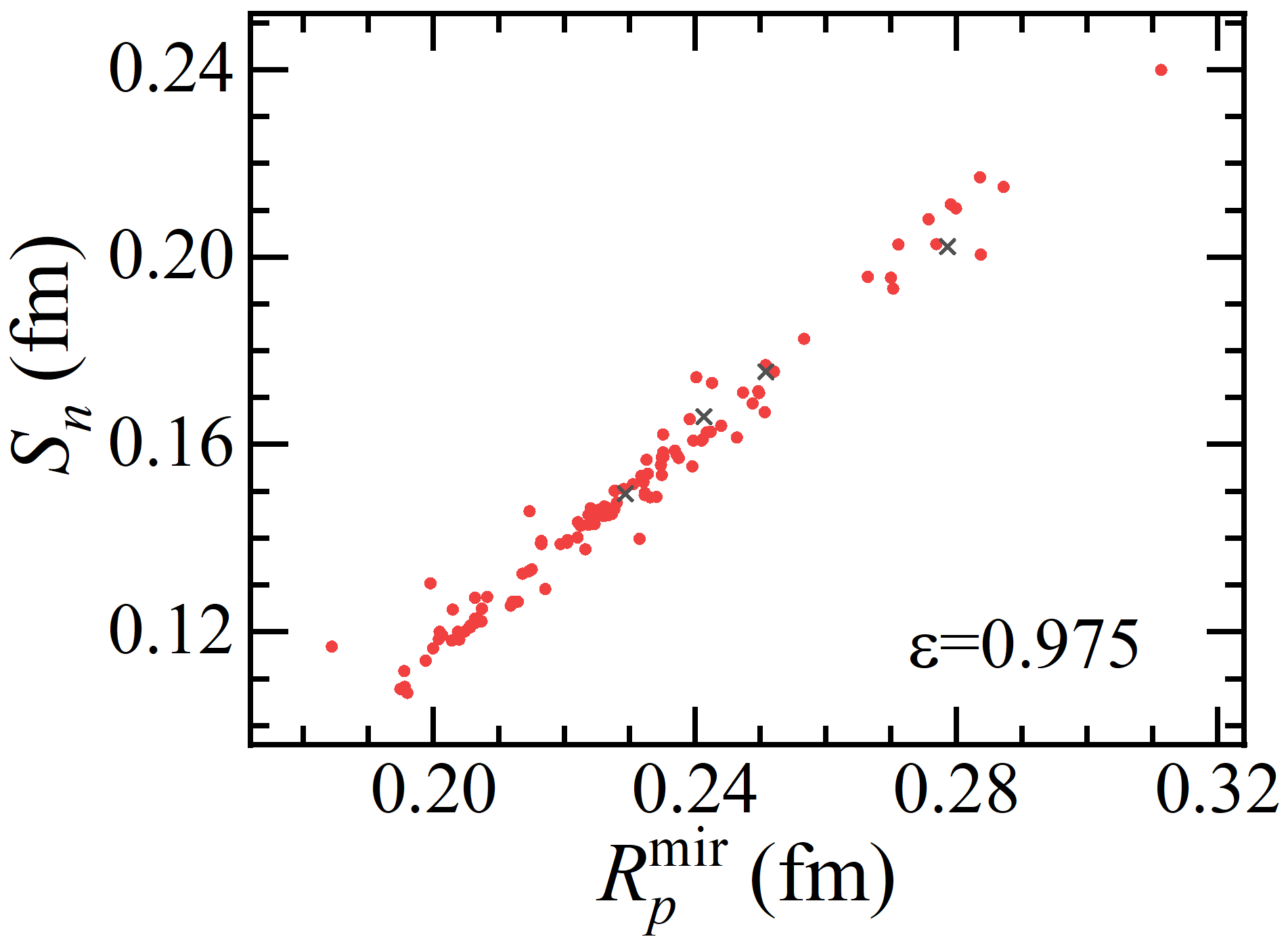}
\caption{Correlation between the difference of the proton radii of mirror nuclei and neutron skin thickness for the \ch{^{48}Ca}--\ch{^{48}Ni} pair.
$\varepsilon$ is the coefficient of determination of linear fit.~\cite{Ding_2023}}
\label{fig.9}
\end{figure}

Mirror nuclei are the nuclei with the same mass number but interchanged proton and neutron number.
Assuming perfect charge symmetry, the neutron radius of a given nucleus is strictly equal to the proton radius of the corresponding mirror nucleus so that the neutron skin thickness can be evaluated through the difference of proton rms radius of the mirror nuclei ($R_{p}^{\rm mir}$).
Although the charge symmetry is slightly broken in reality because of the Coulomb interaction, a linear relationship between the difference of the rms charge radii of mirror nuclei and $|N-Z| \times L$ has been theoretically predicted to still exist~\cite{PhysRevLett.119.122502,PhysRevC.97.014314}.
And it has been shown that the neutron skin thickness is correlated with both $|N-Z| \times L$ and $E_\text{sym}(\rho = 0.10\ \rm fm^{-3})$~\cite{PhysRevLett.119.122502}.
Successive researches are performed to study the effects of other factors on the above suggested relations, such as pairing correlations, low-lying proton continuum, and deformation~\cite{PhysRevC.105.L021301, PhysRevC.107.034319, Ding_2023}.
In Ref.~\cite{Ding_2023}, through the systematical investigations by using axially deformed solution of the Skyrme-Hartree-Fock-Bogoliubov equations with 132 sets of Skyrme interaction parameters, the neutron skin thickness is demonstrated to be in proportion to $R_{p}^{\rm mir}$ even with the consideration of the Coulomb interaction as shown in Fig.~\ref{fig.9}.
Moreover, the pairing effects can enhance the correlation for most mirror pairs while deformation effects may weaken the correlation by comparing with the results of the SHF model.
Based on the experimental value of $R_{p}^{\rm mir}$, it is possible to deduce the neutron skin thickness of neutron-rich nuclei and constrain the range of $L$.
Besides, it is found that the neutron skin thickness of a neutron-rich nucleus is also roughly proportional to $R_{p}^{\rm mir}$ of its isotones and the linearity is stronger with larger $|N-Z|$, which provides an approach to extract the neutron skin thickness of an unstable nucleus even without experimental $R_{p}^{\rm mir}$ data.

In addition, recent studies indicate that the precise measurement of the charge-changing cross section ($\sigma_{cc}$) could be a novel and valid method to extract charge radii of unstable nuclei~\cite{PhysRevC.82.014609, PhysRevLett.107.032502, PhysRevC.89.044602, Li:2016gjs}.
Therefore, the relation between the charge-changing cross section difference of mirror nuclei ($\Delta \sigma_{cc}$) and $L$ or neutron skin thickness has been further investigated~\cite{XU2022137333}.
\begin{figure}[htbp]
\includegraphics[width=0.85\columnwidth]{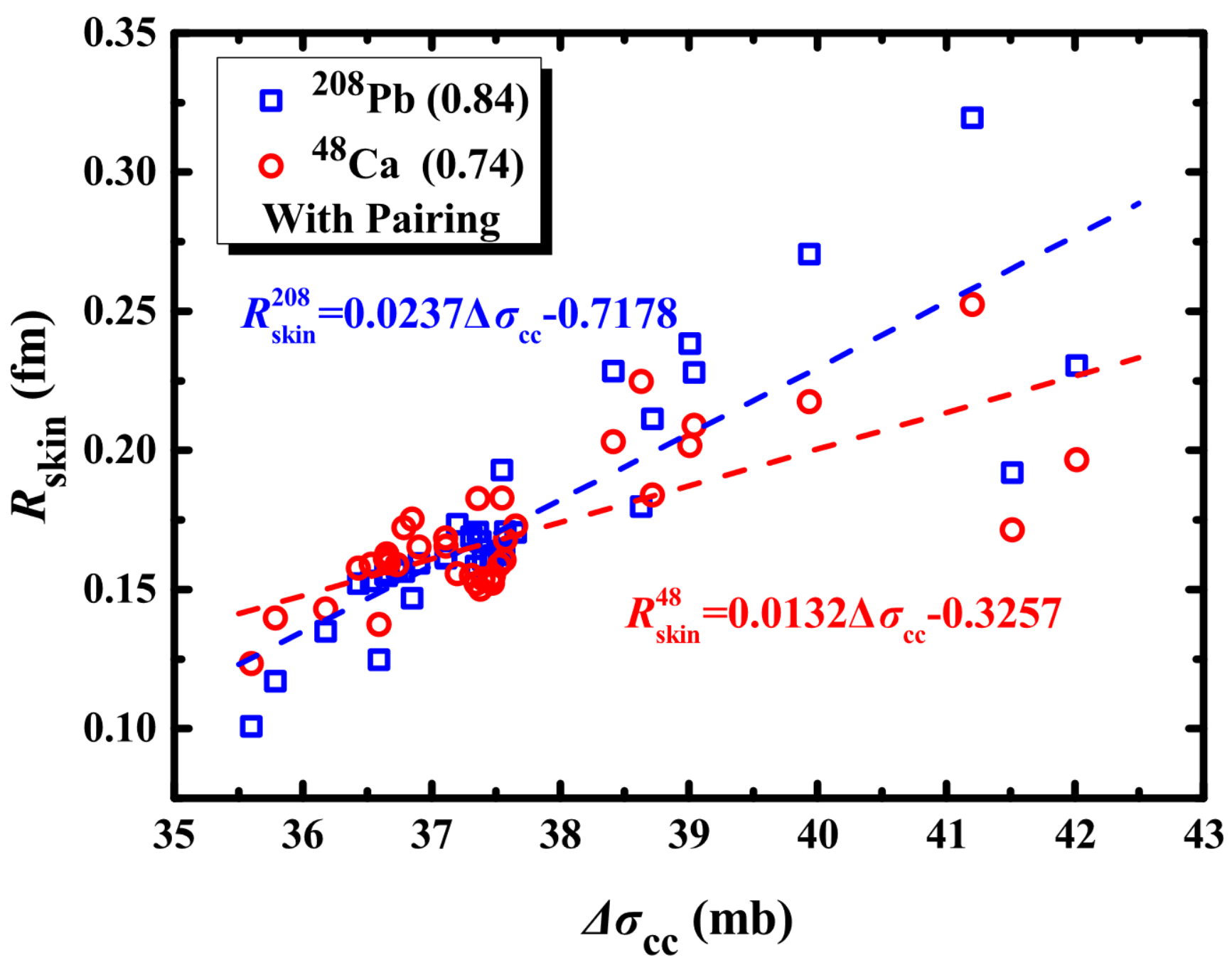}
\caption{Correlation between the difference of the charge-changing cross section of mirror nuclei \ch{^{30}Si}--\ch{^{30}S} and neutron skin thickness of \ch{^{208}Pb} and \ch{^{48}Ca}.
Numbers in parentheses are the coefficients of determination of linear fit.~\cite{XU2022137333}}
\label{fig.10}
\end{figure}
It is found that $\sigma_{cc}$ is perfectly proportional to proton rms radii and $L$ has a good linear correlation with $\Delta \sigma_{cc}$, $R_{p}^{\rm mir}$, or neutron skin thickness for the \ch{^{30}Si}--\ch{^{30}S} pair by using both the SHF theory and covariant (relativistic) density functionals (CDF) together with a Glauber model analysis.
Furthermore, it can be seen from Fig.~\ref{fig.10} that $\Delta \sigma_{cc}$ of \ch{^{30}Si}--\ch{^{30}S} mirror nuclei is sensitive to the neutron skin thickness of both \ch{^{48}Ca} and \ch{^{208}Pb}, meaning that the light mirror nuclei can be used to constrain the EOS which will be implemented more simply with the current experimental facility.
Consequently, the charge-changing cross section difference of mirror nuclei is expected to be an effective surrogate for probing the neutron skin thickness or the density dependence of the symmetry energy.

\subsection{Neutron skin and astrophysical $S$-factor}
The density dependence of symmetry energy and the property of neutron skin are of vital importance in the wide domain ranging from nuclear physics to astrophysics~\cite{PhysRevC.93.044618, PhysRevC.104.024606, PhysRevC.106.044318, WAKASA2021104749, Lattimer_2001, Lattimer:2023rpe}.
Heavy-ion fusion reactions at low incident energies are governed by quantum tunneling through the Coulomb barrier which is formed from the action of repulsive long-range Coulomb interaction and attractive short-range nuclear interaction~\cite{PhysRevC.85.057602, Hagino2020}.
\begin{figure}[htbp]
\includegraphics[width=0.95\columnwidth]{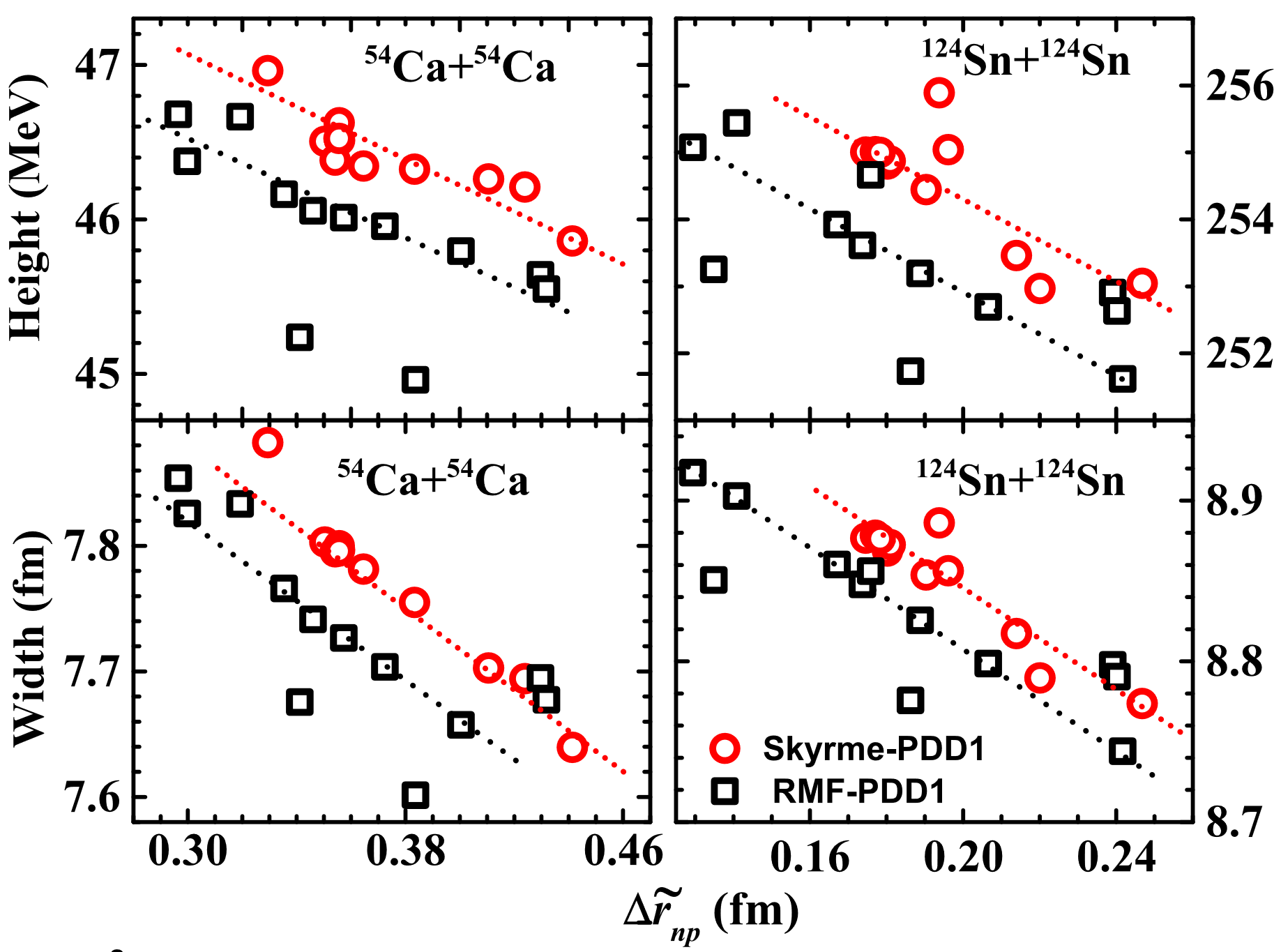}
\caption{Correlation between the barrier parameters (heights and widths) and neutron skin thickness.~\cite{Ghosh:2023edd}}
\label{fig.11}
\end{figure}
\begin{figure}[htbp]
\includegraphics[width=0.95\columnwidth]{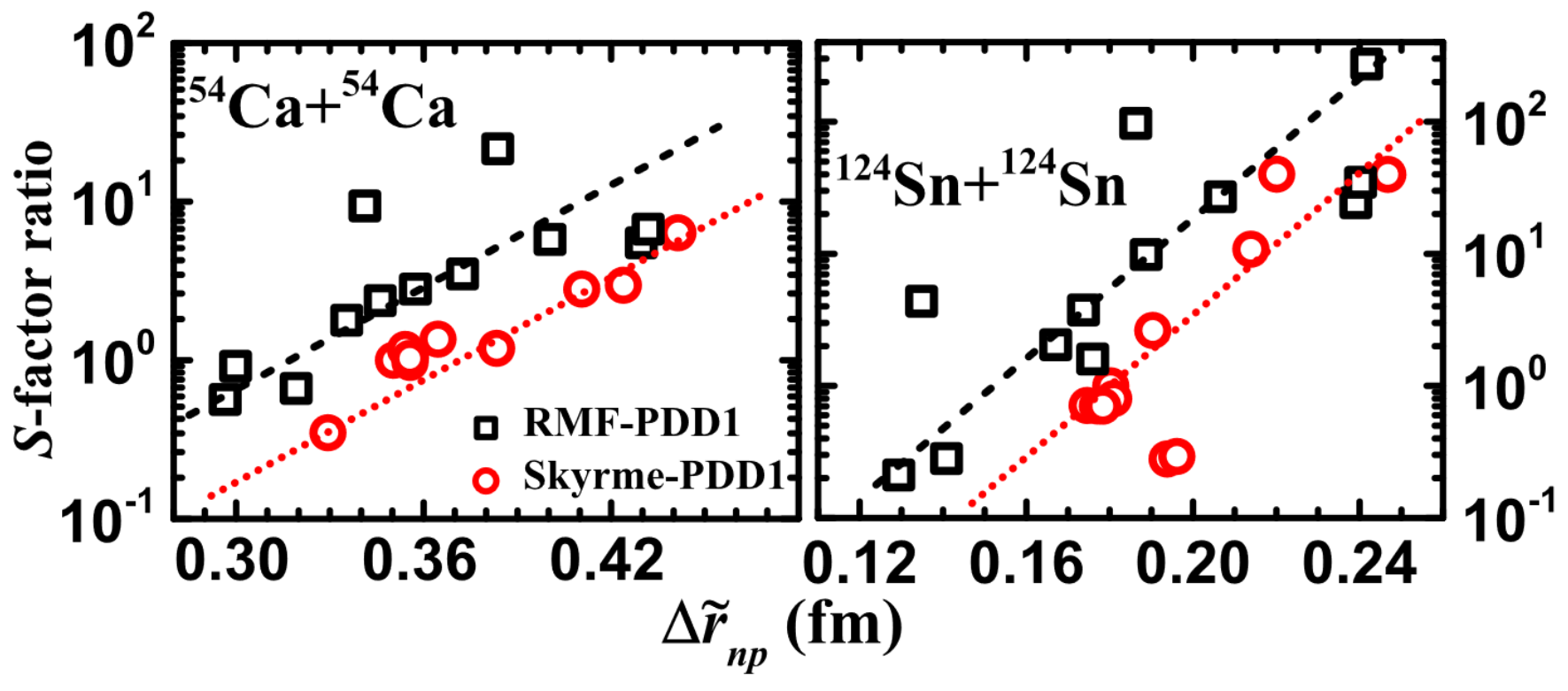}
\caption{Correlation between $S$-factor ratio and neutron skin thickness.~\cite{Ghosh:2023edd}}
\label{fig.12}
\end{figure}
The behavior of sub-barrier fusion reactions is a critical issue for the formation of superheavy elements as well as nuclear astrophysics.
The expression of the cross section can be described as
\begin{equation}\label{eq.7}
\sigma(E)=E^{-1}\exp(-2\pi\eta)S(E),
\end{equation}
where $E$ is the center of mass energy of the reaction system, $\eta$ is the Sommmerfeld parameter, and $S(E)$ is called the astrophysical $S$-factor, which varies weakly with energy and includes nuclear structure effects.
Reference~\cite{Ghosh:2023edd} explores the sensitivity of the neutron skin thickness to astrophysical $S$-factor for heavy-ion fusion cross sections.
The nucleus–nucleus potentials are generated from the double-folding model, where the density distributions of nucleons as key inputs are calculated by different families of non-relativistic and relativistic mean-field models corresponding to a wide range of the neutron skin thickness or $L$.
As shown in Fig.~\ref{fig.11}, the barrier parameters, such as its height and width, are found to decrease with the increasing neutron skin thickness, which leads to the enhancement of cross section and astrophysical $S$-factor up to one or two orders of magnitude.
Moreover, Fig.~\ref{fig.12} clearly illustrates that the $S$-factor is roughly proportional to the neutron skin thickness.
The results from various asymmetric systems manifest that these effects become stronger with increasing proton numbers.
Therefore, it may be probable to determine the neutron skin thickness or $L$ from the precise measurement of sub-barrier fusion cross section or astrophysical $S$-factor.

\section{Neutron skin and high-energy heavy-ion collisions}\label{3}
\begin{figure}[htbp]
\includegraphics[width=0.9\columnwidth]{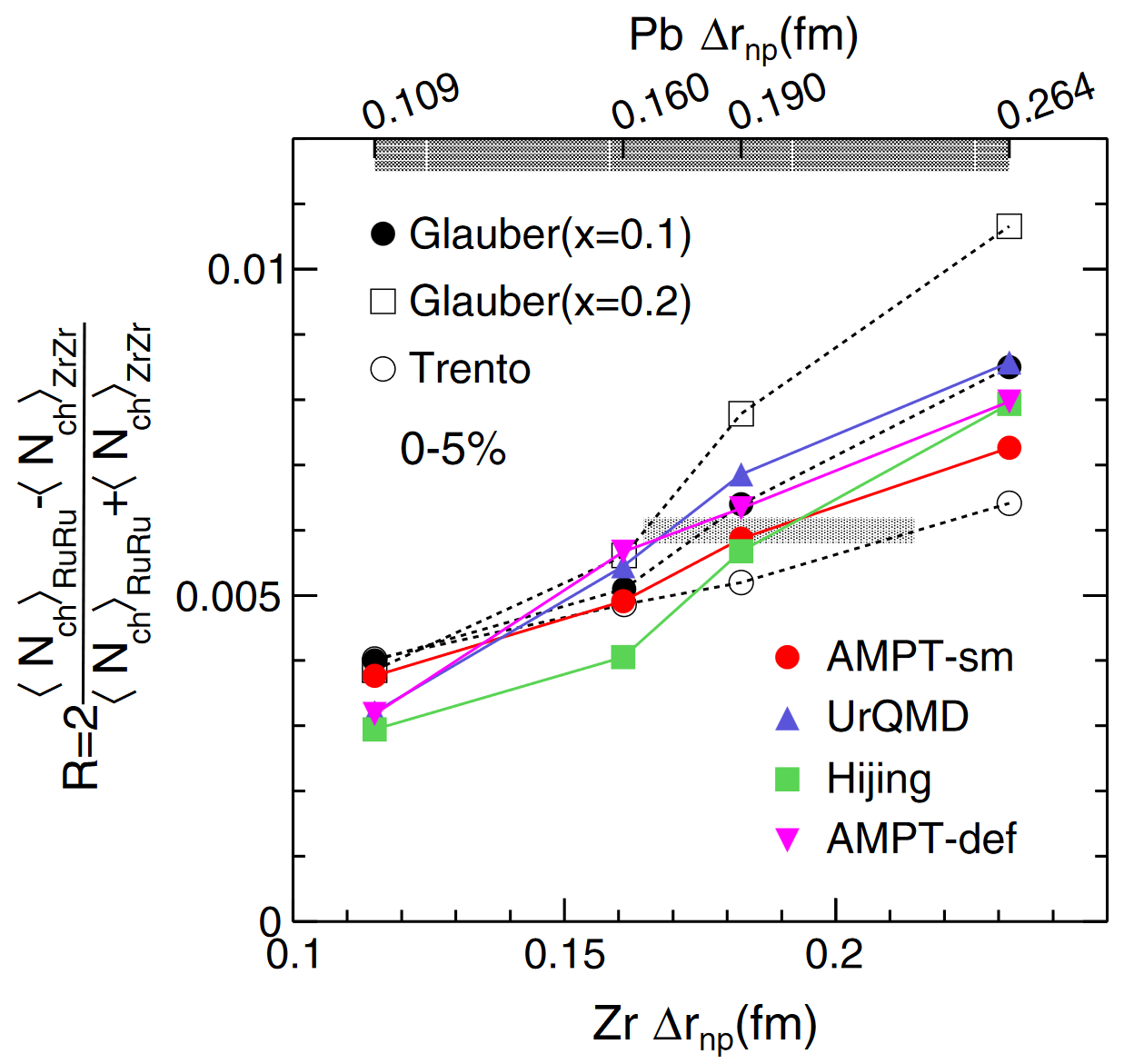}
\caption{Correlation between the relative $\langle N_{\text{ch}} \rangle$ ratio (R) and neutron skin thickness of Zr.~\cite{PhysRevLett.125.222301}}
\label{fig.13}
\end{figure}

Relativistic heavy-ion collisions are crucially important for nuclear physics, which aim at investigating the properties of the hot dense matter known as the quark-gluon plasma~\cite{Chen-PR,Nature23,Sun-NC,ZhangY,SunKJ,LiFP,GaoJH,ZhaoXL,Ma-NST,Ma-CPL,Wang:2023fvy, Ma2023NewTO}.
Recently, extensive attention has been paid to the effects of the exotic nuclear structures, such as neutron skin and deformation, on the final observables in relativistic heavy-ion collisions~\cite{Ma2020, Xu:2022ikx}.
The relevant simulations are carried out based on the isobaric \ch{^{96}_{44}Ru} + \ch{^{96}_{44}Ru} and \ch{^{96}_{40}Zr} + \ch{^{96}_{40}Zr} collisions at ultra-relativistic energies~\cite{PhysRevC.105.014901,PhysRevC.106.014906}. 
Specifically, in Ref.~\cite{PhysRevLett.125.222301}, different neutron skin sizes are calculated via two energy density functional (EDF) theories, namely, the standard SHF and the extended SHF, with various symmetry energy parameters.
With the help of four dynamical models (AMPT-sm, UrQMD, HIJING, AMPT-def) and two static models (TRENTO and Glauber), it is concluded from Fig.~\ref{fig.13} that the charged hadron multiplicity ($N_{\text{ch}}$) difference between the two isobaric collisions is sensitive to the neutron skin thickness, which is weakly model-dependent.
Furthermore, Refs.~\cite{LIU2022137441,PhysRevC.106.034913} present that the neutron skin thickness is closely related to not only the yield ratio of free spectator neutrons produced in high-energy isobaric collisions but also the yield ratio of free spectator neutrons to protons in a single collision system.
In addition, the ratios between isobar collisions of the mean transverse momenta~\cite{PhysRevC.108.L011902} and the net charge multiplicities~\cite{PhysRevC.105.L011901} are demonstrated to be reliable probes of the neutron skin difference between the isobar nuclei.
In consequence, these investigations indicate the possibility of utilizing the observables in high-energy heavy-ion collisions to probe the neutron skin thickness and further give stringent constraints on the nuclear symmetry energy.

\begin{figure}[htbp]
\includegraphics[width=0.9\columnwidth]{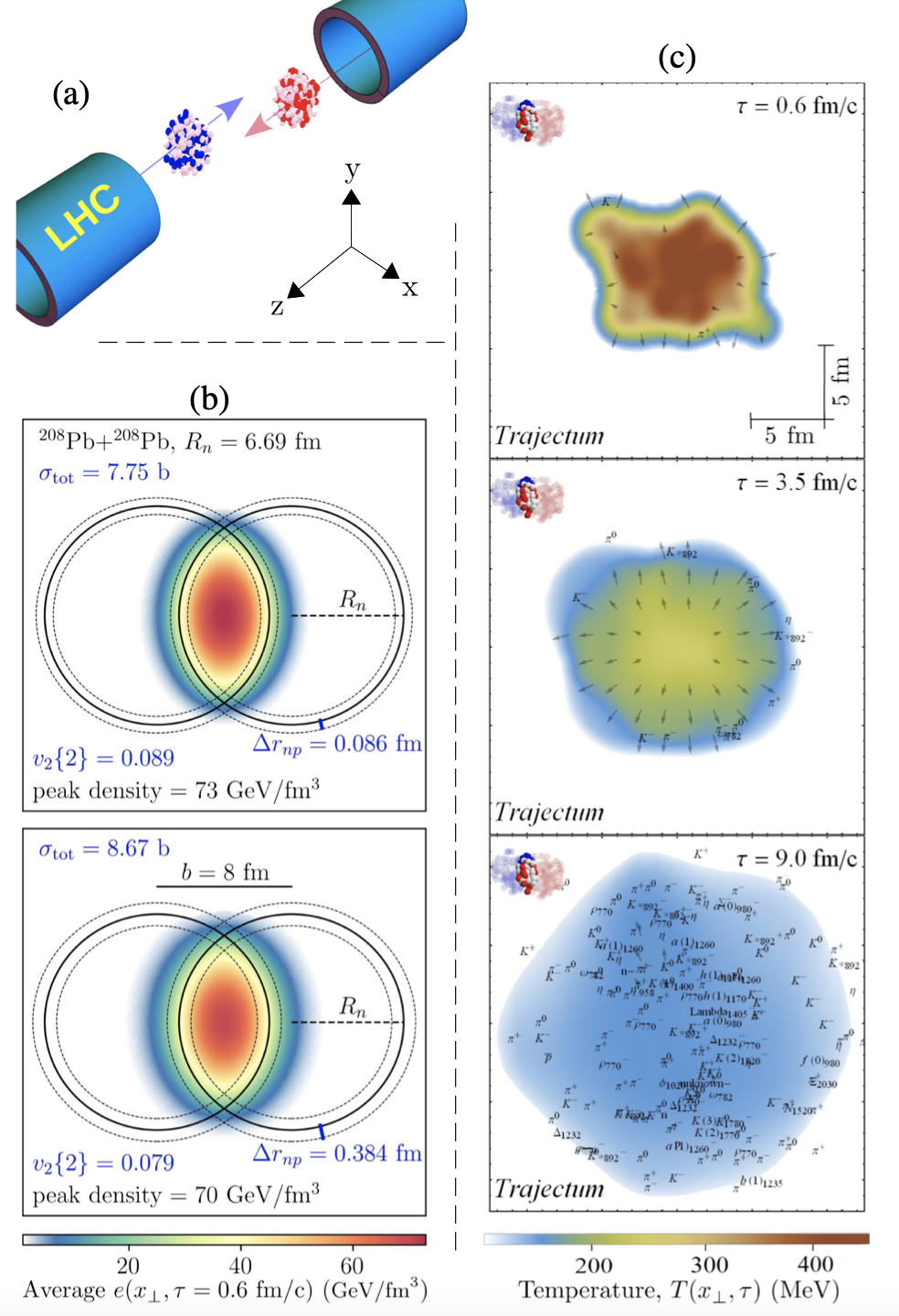}
\caption{Neutron skin and collective flow in relativistic heavy-ion collisions. (a) Two ions collide with impact parameter b=8 fm. 
(b) The collision deposits energy in the interaction region depending on the extent of the neutron skin of the $^{208}$Pb nuclei. Here $\Delta r_{np}$ = 0.086 (top) and $\Delta r_{np}$ = 0.384 fm (bottom) are used. 
(c) A single QGP is evolving hydrodynamically and  converted into particles 
as it cools, while expanding both in $z$ and in the transverse plane, and finally leads to azimuthal anisotropies, i.e. collective flow.~\cite{PhysRevLett.131.202302} 
}
\label{fig.14}
\end{figure}

Recent, an attempt for determination of the neutron skin of $^{208}$Pb from ultrarelativistic nuclear collisions has been carried out. As a typical heavy nucleus, $^{208}$Pb has been the target of many dedicated efforts due to well established nuclear structure information. In Ref.~\cite{PhysRevLett.131.202302}, they analyzed the measurement results of particle distributions and their collective flow in $^{208}$Pb+$^{208}$Pb collisions at ultrarelativistic energy performed at the Large Hadron Collider to extract the neutron skin thickness. Fig.~\ref{fig.14} shows an ultrarelativistic heavy-ion collision in the lab frame (a), where collisions deposit energy density in the area of overlap (transverse plane) that is perpendicular to the beam direction (b).  A QGP fluid  evolves within the hydrodynamic model of heavy-ion collisions as shown in  Fig.~\ref{fig.14}(c) until the confinement crossover is reached, then the cooled  QGP fluid is  converted into a gas of QCD resonance states that can further rescatter or decay to stable particles. 
By means of global analysis tools, they infer a neutron skin thickness $\Delta r_{np} = 0.217 \pm 0.058$ fm, consistent with nuclear theory predictions, and competitive in accuracy with a recent determination from parity-violating asymmetries in polarized electron scattering. In this way, a new experimental method was proposed to systematically measure neutron distributions in the ground state of atomic nuclei.

In another work, the STAR Collaboration presented a method  by the photoproduction of vector mesons in ultraperipheral nucleus-nucleus collisions to infer the average gluon density in the colliding nuclei, and hence the neutron skins \cite{SciAdvance,Ma2023NewTO}.
When two relativistic heavy nuclei pass through at a distance of a few nuclear radii, the photon from one nucleus may interact through a virtual quark-antiquark pair with gluons from the other nucleus, forming a short-lived vector meson, such as $\rho_0$. In this experiment, the polarization was used in diffractive photoproduction to observe a unique spin interference pattern in the angular distribution of ${\rho_0 \rightarrow \pi^+ \pi^-}$ decays. The observed interference can be explained by an overlap of two wave functions at a distance an order of magnitude larger than the  $\rho_0$ travel distance within its lifetime. The strong-interaction nuclear radii were extracted from these diffractive interactions and found to be 6.53 $\pm$ 0.06 fm ($^{197}$Au) and 7.29 $\pm$ 0.08 fm ($^{238}$U), larger than the nuclear charge radii. The observable is demonstrated to be sensitive to the nuclear geometry and quantum interference of nonidentical particles.

The progress along this direction using relativistic heavy ion collision is quite promising. Traditionally, neutron skin effects, as one of nuclear structure effects, are believed only playing an important role in low energy nuclear physics. Due to the ultra-fast process in high energy interaction, transient images of atomic nuclei can be photographed. Actually, an impressive work demonstrates that it is possible to obtain deformation parameters of atomic nuclei, traditionally deduced from low-energy nuclear physics experiments, from the momentum distribution of particles produced in high-energy nuclear collisions \cite{arxiv_STAR}. Thus, a bridge between high energy and low energy heavy ion physics is setting up.

\section{Experimental methods for neutron halo and neutron skin} \label{4}
The emergence of new phenomena and physics in unstable nuclei far from the stability line poses a challenge to the traditional nuclear theory.
In order to have an explicit knowledge of them, it is necessary to generate and separate the radioactive nuclear beams with high quality and high intensity through the accelerator facilities, which is the basis of experimental studies with RNB.
This section will introduce the experimental methods of producing RNB and probing the structure of neutron halo and neutron skin. 

\subsection{Production methods for radioactive nuclear beam}
\begin{figure}[htbp]
\includegraphics[width=0.8\columnwidth]{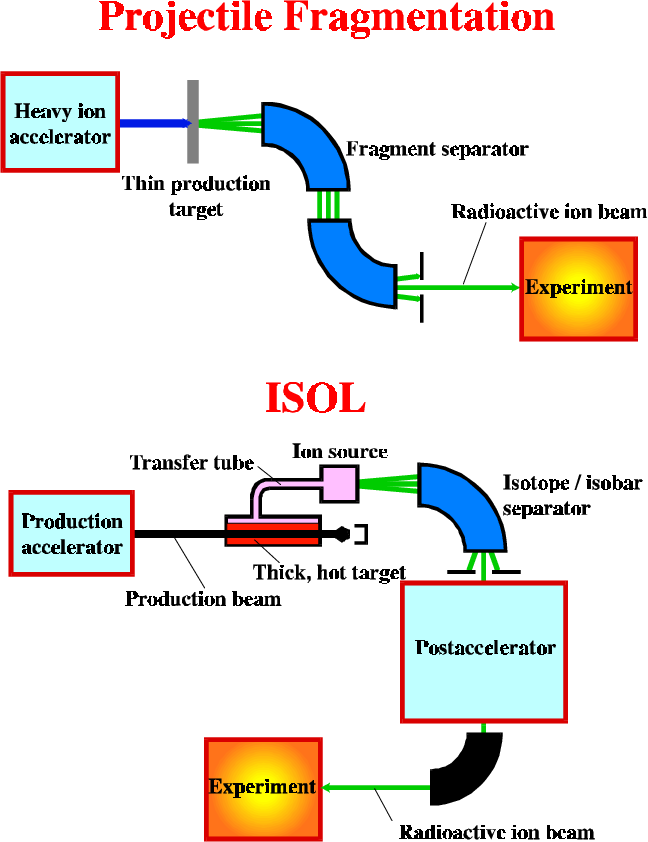}
\caption{Schematic view for the two types of radioactive nuclear beam facilities.~\cite{doi:10.1080}}
\label{fig.15}
\end{figure}

The unstable nuclei in experimental studies are principally produced by facilities of RNB.
Various types of RNB facilities have been built and put into operation in major national laboratories of nuclear physics in the world, including LISE at GANIL (France), A1200 at NCSL (USA), RIPS at RIKEN (Japan), FRS at GSI (Germany), and RIBLL at HIRFL (China)\cite{MA2021103911}.
In particular, the High Intensity heavy-ion Accelerator Facility 
(HIAF) is under construction at the Institute of Modern Physics (IMP), Chinese Academy of Sciences, which can provide high-intensity heavy-ion beams.
It is hopeful to offer pioneering conditions for identifying new nuclides, expanding the nuclear landscape, and studying the exotic phenomena and physics in nuclei far away from the stability line.
So far, there are two basic approaches to generate RNB: projectile fragmentation (PF) and isotope separation on-line (ISOL).
The schematic is illustrated in Fig.~\ref{fig.15}~\cite{doi:10.1080}.

For the PF method, the primary accelerator directs a heavy-ion beam with certain energy on a thin production target, which produces a variety of nuclear fragments with different charge and mass numbers.
Subsequently, these fragments are selected by the electromagnetic separator through the conditions of mass, charge, or momentum to get the secondary radioactive ion beam.
This method makes it possible to produce the unstable nuclei with a very short lifetime in $\mu$s range, including many secondary beams close to the neutron- and proton-drip lines.
The maximum energy of the secondary beams is close to that of the projectile.
The energy can be reduced by using an energy degrader when lower energy is required by the experiment.
Most of the second generation of RNB devices currently in use are PF-type on account of the lower technical difficulty, lower investment, and higher efficiency, such as LISE, A1200, RIPS, FRS, and RIBLL as mentioned above.
For the ISOL method, radioactive nuclei are produced through multifarious nuclear reactions by bombarding high-intensity light-particle beams with intermediate to high energies on a thick target.
Most or even all of the energy of the beam is lost in the target.
The produced nuclei will be ionized on line and then selected by an isotope/isobar separator into the nuclear species of interest with a low energy. Then the beam is eventually injected into a post-accelerator to be accelerated to an energy appropriate for the experimental terminal which can be around and above the Coulomb barrier (10-20 MeV/nucleon).
This method is conducive to generate secondary radioactive beams with high intensity up to $10^{11}$--$10^{12}$ pps (particle per second) and high quality, while with the lifetime longer than 500 ms because the ISOL requires some time, making it difficult to obtain short-lived nuclei.
Representative ISOL-type facilities are ISOLDE at CERN, SPIRAL at GANIL, TRIUMF at Canada, etc.
These two different technologies have their own advantages and are complementary technically, so they can be selected according to actual needs.
More details about RNB facilities and methods can be found in Refs.~\cite{doi:10.1080, Blumenfeld_2013, MA2021103911}.

\subsection{Experimental methods for neutron halo}
Since the discovery of neutron-halo structure in \ch{^{11}Li} by I. Tanihata~\cite{PhysRevLett.55.2676, TANIHATA1985380}, a series of observables sensitive to halo structure have been proposed, and the corresponding experimental methods have been developed.
Compared to stable nuclei, halo nuclei exhibit different properties in many aspects.
The two significant characteristics are larger matter radii and narrower width of the momentum distribution of valence nucleons.
Direct measurement of nuclear reactions is the primary experimental method to study the halo structure.
However, this method is only suitable for nuclear ground states or long-lived excited states because the measurement is performed by bombarding the radioactive beam on targets.
The cross section of reactions and momentum distribution of fragments are two classical quantities which can be easily extracted in the direct measurement of nuclear reactions and are sensitive to exotic nuclear structure, especially the halo structure.
Most of the current experiments for halo nuclei are performed by measuring these two quantities.

\subsubsection{Reaction cross section}
In experiments, the charge radii of the long-lived nuclei can be measured by electron scattering, proton scattering, isotope shift, etc.
However, the nuclear matter radii is commonly determined by measuring the total cross section of nuclear reactions.
The total nuclear reaction cross section is the sum of probabilities of various reactions in nuclei-nuclei collisions, which depends on several factors, such as the radius of the projectile and target nucleus and the collision energy.
At high energies, the total nuclear reaction cross section ($\sigma_{R}$) is in direct proportion to the square of the sum of the radii of the projectile ($R_{P}$) and the target ($R_{T}$) approximately, which can be simply expressed as
\begin{equation}\label{eq.8}
\sigma_{R}=\pi(R_{P}+R_{T})^2.
\end{equation}
It is noted that $R_{T}$ is regarded as a known quantity because the target is a stable nucleus.
So if $\sigma_{R}$ is measured in experiments, $R_{P}$ or the density distribution of the projectile can be deduced by Eq.~\ref{eq.8}, which is the principle of measuring the nuclear matter radius.

Nevertheless, in reality $\sigma_{R}$ is dependent on energy which can be well described by different kinds of theoretical models.
In theory, the commonly used methods for calculating $\sigma_{R}$ include Kox parameterized formula~\cite{PhysRevC.35.1678}, Shen parameterized formula~\cite{SHEN1989130}, Glauber model based on semi-classical approximation of quantum mechanics~\cite{PhysRevC.41.1610}, and microscopic transport theory such as Boltzmann-Uehling-Uhlenbeck (BUU)~\cite{MA1993386, PhysRevC.48.850}.
The density distribution of nuclear matter is an indispensable input in both Glauber and transport model.
Therefore, by using these models to fit the experimental data of $\sigma_{R}$, the density distribution of nuclear matter can be extracted, and then the rms radius of nuclear matter can be obtained.
These methods for calculating $\sigma_{R}$ have been widely used to study the effects of halo or skin structure so as to explore its formation reasons and characteristics.

In experiments, $\sigma_{R}$ is usually measured by the transmission method~\cite{PhysRevC.69.034613, PhysRevC.76.031601}.
When the beam passes through the target, the number of incident particles ($N_{0}$) is correlated with the number of outgoing particles without a reaction ($N_{1}$), the target thickness ($t$), and $\sigma_{R}$, which can be described as:
\begin{equation}\label{eq.9}
N_{1}=N_{0}\exp(-\sigma_{R}t).
\end{equation}
This expression can be transformed into:
\begin{equation}\label{eq.10}
\sigma_{R}=-\ln(N_{1}/N_{0})/t,
\end{equation}
Hence, under the condition of the known $t$, $\sigma_{R}$ can be deduced by measuring $N_{0}$ and $N_{1}$ in experiments, which can be realized by setting detectors in front of and behind the reaction target.
However, this is just the ideal situation.
In practice, both the detection efficiency of particles and the transfer efficiency of the beam in devices cannot reach 100$\%$.
Moreover, the incident beam may react with all detectors besides the target.
Consequently, these effects should be corrected by adding an empty-target measurement as described in Ref~\cite{PhysRevC.69.034613}.

\subsubsection{Fragment momentum distribution}
Another quantity that can be measured directly in nuclear reactions is the momentum distribution of the fragments through nucleon removal or breakup reactions, which can be either the core or the valence nucleon(s).
The nuclei with halo or exotic structures are generally considered to consist of a core plus valence nucleons.
The uncertainty principle qualitatively explains why the momentum distribution of fragments is sensitive to the halo structure.
The narrower the momentum distribution width is, the smaller the intrinsic momentum fluctuation between the core and valence nucleons is in the projectile rest system.
According to the uncertainty principle, it signifies that the core and valence nucleons form a relatively loose structure in coordinate space, that is, the spatial distribution of the valence nucleons is diffused away from the core which is exactly the essential characteristic of halo structure.
Additionally, from the perspective of the representation transformation in quantum mechanics, the wave functions in coordinate and momentum representations are intimately related to each other by the Fourier transform.
A nucleon with a wider density distribution in the coordinate space appears as a narrower distribution in the momentum space.
Moreover, relative to the value of the wave function in momentum space at zero momentum, the full width at half maximum (FWHM) of momentum distribution is more sensitive to the halo structure.

In the projectile rest frame, the core part of the nucleus has exactly the same momentum as the valence nucleons but with opposite direction.
Therefore, the momentum distribution of the core can be measured to obtain the momentum distribution of the valence nucleons in experiments.
Especially for some neutron-rich nuclei, their valence nucleons are neutrons which is difficult to be measured.
Thus, it is advisable to measure the momentum distribution of the core after removal of the valence nucleons.
Instead of the transverse momentum, the measurement of the longitudinal momentum has less broadening effects introduced by Coulomb deflection and multiple scattering in the thick breakup targets, and its FWHM can effectively reflect the internal structure of the projectile nucleus~\cite{PhysRevC.51.3116}.

Some measuring methods have been developed to extract the momentum distribution of fragments in the process of bombarding projectiles on targets.
Among them, the classical approaches are magnetic spectrometers, energy-loss spectrometers, and direct time-of-flight (TOF).
A summary of magnetic spectrometers and energy-loss spectrometers methods can be referred to Ref.~\cite{ORR1997155}.
The TOF analysis is that the velocity and momentum of a particle are determined by the time of a known flight distance~\cite{PhysRevC.48.R1484, PhysRevC.69.034613}.
Compared with the measurements by magnetic spectrometers or energy-loss spectrometers, the direct TOF technique has the following advantages.
Firstly, it has a wider longitudinal momentum acceptance, and a broader range of the momentum spectrum which weakens the influences of distortion or dissociation on the spectrum width in the process of fitting.
However, there still exists momentum distribution data far from the peak position.
Secondly, the direct TOF method allows for simultaneous measurements of the momentum distributions of different nuclear species produced by the same projectile, as well as the momentum distributions of fragments produced by different radioactive secondary beams.
Consequently, it enables to obtain multiple groups of data at the same time in a single experiment, and realize simultaneous measurements of the total cross section of nuclear reactions and momentum distributions of fragments.
It can also carry out systematic researches on different reaction systems in one experiment, which is of great significance for improving the experimental efficiency.

\subsubsection{Experiments at RIPS}
\begin{figure}[htbp]
\includegraphics[width=0.95\columnwidth]{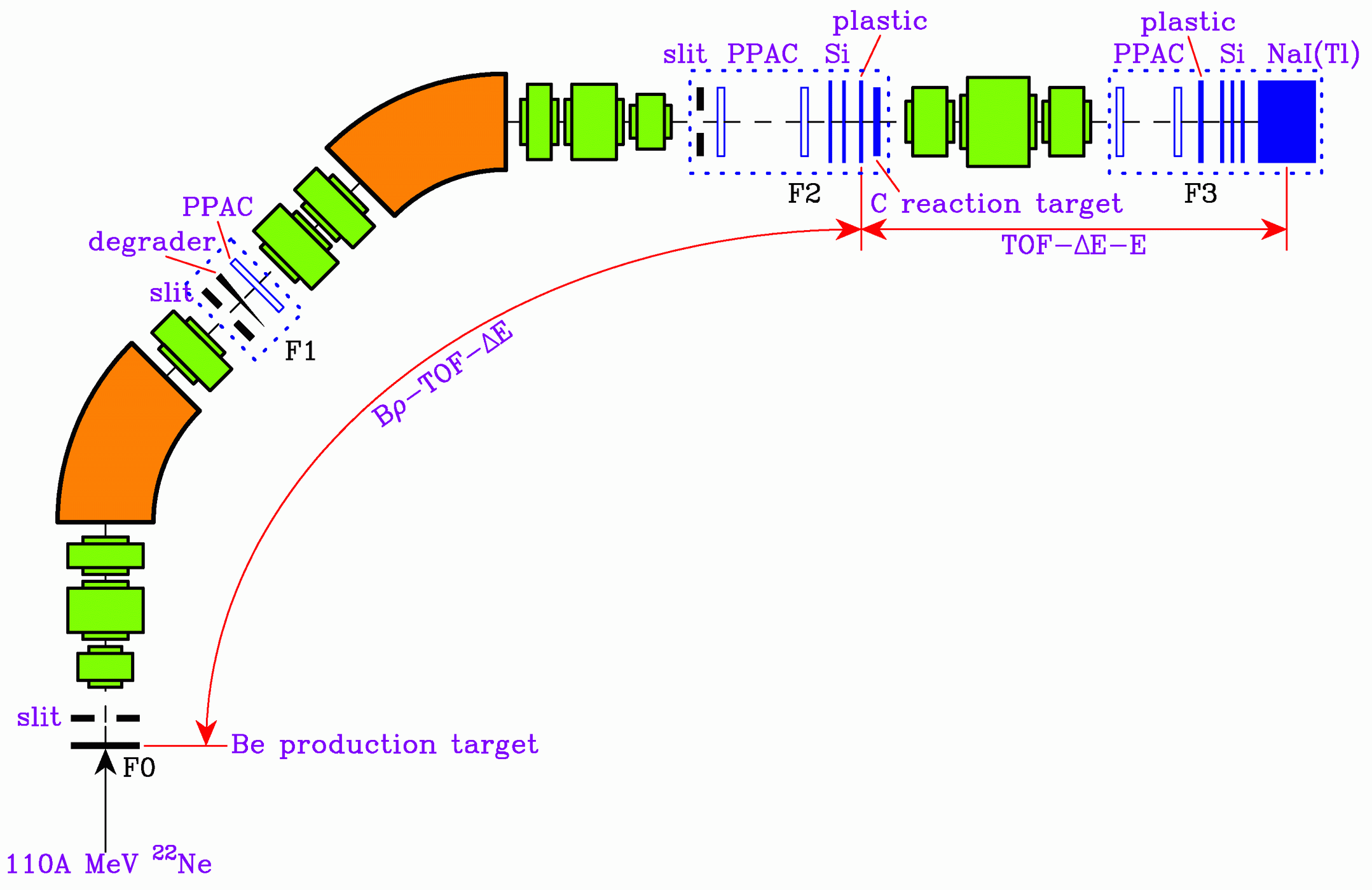}
\caption{Experimental setup at the fragment separator RIPS.~\cite{PhysRevC.69.034613}}
\label{fig.16}
\end{figure}

In the experiments performed at the RIKEN projectile fragment separator (RIPS) in the RIKEN Ring Cyclotron Facility, the longitudinal momentum distribution of fragments is measured by a direct TOF technique and the total reaction cross section is determined by a transmission method to investigate the exotic structure of the neutron-rich nucleus \ch{^{15}C}~\cite{PhysRevC.69.034613} and the proton-rich nucleus \ch{^{23}Al}~\cite{PhysRevC.76.031601}.
The experimental setup for \ch{^{15}C} is shown in Fig.~\ref{fig.16}.

In the F0 chamber, the 110 MeV/nucleon \ch{^{22}Ne} primary beam was bombarded on a Be target to produce the secondary beams of \ch{^{14, 15}C} through projectile fragmentation reactions.
In the dispersive focus plane F1, an Al wedge-shaped degrader was installed to separate particles by energy-loss method.
A delay-line readout parallel plate avalanche counter (PPAC) was placed to measure the beam position.
Then the secondary beam was directed onto the achromatic focus F2.
Two charge-division readout PPACs were used to determine the beam position and angle.
Two Si detectors were used to measure the energy losses ($\Delta E$), and an ultrafast plastic scintillator was placed before the C reaction target to measure the TOF from the PPAC at F1.
The particle identification before the reaction target was done by the $B\rho$--$\Delta E$--TOF method.
Subsequently, a quadrupole triplet was used to transport and focus the beam onto F3.
Two delay-line readout PPACs were used to monitor the beam size and emittance angle.
Another plastic scintillator gave a stop signal of the TOF from F2 to F3. 
Three Si detectors was used to measure the energy loss ($\Delta E$), and a NaI(Tl) detector was placed to measure the total energy ($E$).
The particles were identified by the TOF--$\Delta E$--$E$ method.
The longitudinal momentum distributions of the fragment after the removal of nucleons were determined from the TOF between the two plastic scintillators at F2 and F3.
And the position information from the PPAC at F1 was used to derive the incident momentum of the beam.
In addition, the reaction cross section was extracted with and without a reaction target by a transmission-type method via Eq.~\ref{eq.10}.
The experimental data is reproduced by the calculations by the few-body Glauber model to extract useful information on the density distribution of nuclear matter, the nuclear radius, the orbit and density distribution of valence nucleons for the exotic nuclei far from the stability line.

\subsection{Experimental methods for neutron skin}
Actually, the total nuclear matter radius could be determined from the total nuclear reaction cross section.
In order to obtain the neutron skin thickness, it is necessary to know the rms radii of both protons and neutrons.
Among different cross sections in nuclear reactions, the charge-changing cross section is directly related to the proton radius.
Therefore, the simultaneous determinations of the matter radii from the total nuclear reaction cross section and the proton radii from the charge-changing cross section in experiments provide a way to obtain the neutron skin thickness. 
In addition, all of the observables discussed above, which exhibit strong dependence on the neutron skin thickness, could be used as possible probes to determine the neutron skin thickness from experimental measurements.
Measuring different observables need different experimental devices and setups.
For example, the identification of the projectile and the isotopes of the projectile is vital for measuring neutron removal cross sections. 
The yield ratio of light particles is obtained by measuring neutrons, protons, and charged particles.
The measurement of the photon yield ratio needs the detection of $\gamma$ rays.
More importantly, if those observables are utilized in practice, more systematic studies are required to clarify the sensitivity of each observable to the neutron skin thickness.
Only if the accuracy of the extracted neutron skin thickness meets the expectation, the observable will be an ideal experimental probe.
For example, it is estimated that if the uncertainty of neutron-to-proton yield ratio reaches 5$\%$, the estimated error of the neutron skin thickness could be about 0.1 fm~\cite{SUN2010396}.
Besides intermediate-energy heavy-ion collisions, some probes are expected to be proposed to detect the neutron skin thickness in high-energy reactions, which is limited to merely having the ability to study the neutron skin of some stable nuclei in current studies.
If collisions of unstable nuclei at high-energies can be carried out in the future, some sensitive probes in high-energy nuclear collisions could also be used to study the structure of neutron skin.

In addition to the above methods, the measurement of the parity-violating asymmetry in the elastic scattering of
polarized electrons, carried out at the
Thomas Jefferson National Accelerator Facility, has received much attention.
This experiment provides a more precise electroweak determination of the neutron skin thickness.
The neutron skin thicknesses of \ch{^{208}Pb} ($\Delta R_{np}^{208}$) and \ch{^{48}Ca} ($\Delta R_{np}^{48}$) have been measured.
The first measurement for \ch{^{208}Pb} (PREX-1) inferred $\Delta R_{np}^{208}$(PREX1) $=0.33^{+0.16}_{-0.18}$ fm~\cite{PhysRevLett.108.112502}, and then a new result (PREX-2) with greatly improved precision reported $\Delta R_{np}^{208}$(PREX2) $=0.283\pm0.071$ fm~\cite{PhysRevLett.126.172502}.
In addition, the latest measurement for \ch{^{48}Ca} (CREX) concluded the $\Delta R_{np}^{48}$(CREX) is 0.121 $\pm$ 0.026(exp) $\pm$ 0.024(model) fm~\cite{PhysRevLett.129.042501}.
Within a specific class of relativistic EDFs, a value of $L=106\pm37$ MeV was constrained by $\Delta R_{np}^{208}$(PREX2)~\cite{PhysRevLett.126.172503}, which indicates a stiff EOS and is fairly larger than the previous theoretical and experimental results.
A re-analysis of the PREX data by means of different families of nuclear EDFs suggests $L=54\pm8$ MeV, implying a relatively soft EOS.
Thus, the predicted value of the $L$ parameter based on the same PREX-2 experiment is still under discussion.
Very recently, by using 207 non-relativistic and relativistic mean-field models, the values of $L$ were deduced in the range of 76--165 MeV from $\Delta R_{np}^{208}$(PREX2) and 0--51 MeV from $\Delta R_{np}^{48}$(CREX), which indicates there is no overlap between the two ranges of $L$ making it a big problem to be solved~\cite{TAGAMI2022106037}.
With the Bayesian inference method and the Skyrme EDF, the combined analysis of the CREX and PREX-2 data results in a softer symmetry energy and thinner neutron skin thickness, i.e., $L=17.1^{+39.3}_{-36.0}$ MeV, $\Delta R_{np}^{208}=0.136^{+0.059}_{-0.056}$ fm, and $\Delta R_{np}^{48}=0.150^{+0.031}_{-0.030}$ fm at 90$\%$ confidence level~\cite{PhysRevC.108.024317}.
As a result, the inconsistent results call for further theoretical and experimental investigations.

\section{Discussion and Summary}\label{5}
The structure of neutron skin is an interesting phenomenon for neutron-rich nuclei far from the stability line.
In comparison with the stable nuclei which have similar density distributions of neutrons and protons, the size of neutron skin is closely related to the single-particle orbit and isospin evolution of the shell structure.
The systematic study of the neutron skin has great importance for a deep understanding of the essence of nuclear forces.
In addition, the structure of neutron skin is closely related to the properties of the EOS of asymmetric nuclear matter and neutron stars.
Due to the low experimental accuracy of the neutron radius, the explorations of effective experimental observables that are sensitive to the neutron skin thickness have become one of the hot topics in nuclear physics.

Since the total reaction cross section $\sigma_R$ is one of the most important observables to determine the size of nuclei, W. Q. Shen et al. have studied $\sigma_R$ for heavy ion collisions and its relation with isospin systematically~\cite{SHEN1989130}. By analyzing the variation of $\sigma_R$ with energy from low ($<10$ MeV/nucleon) to intermediate energy for a wide variety of projectile-target combinations, 
an unified parametrized formula for $\sigma_R$ was proposed which can reproduce 
data from low energy to intermediate energy well. 
The effects of neutron skin, ground-state deformation and the surface diffuseness of neutron distribution on $\sigma_R$ were studied via a modified microscopic model in which the difference between neutron and proton distributions in the nucleus are taken into account. 
Assuming that the surface diffuseness of the neutron distribution increases linearly with ($N-Z$), the modified microscopic model can reproduce the dependence of $\sigma_R$ with the neutron excess quite well. The possibility of extracting the neutron skin thickness through the neutron-rich flow in damped collision for neutron-rich reaction systems was also discussed. These results are very important for further studies of nuclear  size and neutron skin thickness.

Recent years, based on various theoretical model, some quantities in nucleus-nucleus collisions or nuclear structure are demonstrated to have strong correlations with the neutron skin thickness, making it possible to extract information on the neutron skin and further constrain the properties of the EOS and neutron stars. In particular, relativistic heavy-ion collisions provide a new venue to investigate the nuclear structure, such as neutron skin, deformation as well as clustering structure. The study on structures and collisions of atomic nuclei is becoming more intersectional and integrated. Of course, further studies are essential on not only the precise experimental determination of the proposed observables but also a search for cleaner probes of the neutron skin with higher sensitivity.

\bibliography{neutronskin_ref}
\end{document}